\documentclass[a4paper,12pt]{article}
\pdfoutput=1

\usepackage{jheppub}

\usepackage[left]{lineno}

\usepackage{graphicx}
\usepackage{amssymb}
\usepackage{epstopdf}
\usepackage{bm}
\usepackage{bbm}
\usepackage{color}
\usepackage[utf8]{inputenc}
\usepackage[T1]{fontenc}
\usepackage[english]{babel}
\usepackage[dvipsnames]{xcolor}
\usepackage{color}
\usepackage{capt-of}
\usepackage{pstricks}
\usepackage{slashed}
\usepackage{booktabs}
\usepackage{axodraw4j}
\usepackage{tikz}
\usepackage{pgfplots}
\usetikzlibrary{shapes,snakes}
\usetikzlibrary{arrows, decorations.markings}
\pgfplotsset{/pgf/number format/use comma,compat=newest}
\pgfrealjobname{\jobname}

\tikzstyle{vecArrow} = [thick, decoration={markings,mark=at position
   1 with {\arrow[semithick]{open triangle 60}}},
   double distance=1.4pt, shorten >= 5.5pt,
   preaction = {decorate},
   postaction = {draw,line width=1.4pt, white,shorten >= 4.5pt}]
\tikzstyle{innerWhite} = [semithick, white,line width=1.4pt, shorten >= 4.5pt]

\usepackage{bm,amsmath,amssymb}

\DeclareMathOperator{\sech}{sech}

\long\def\comment#1{ }

\newcommand{\be}{\begin{eqnarray}}
\newcommand{\ee}{\end{eqnarray}}
\newcommand{\nn}{\nonumber}

\newcommand{\rme}{{\rm e}}

\def\q{{\bm q}}
\def\p{{\bm p}}

\def\l{{\boldsymbol l}}

\def\x{{\boldsymbol x}}
\def\y{{\boldsymbol y}}

\def\r{{\boldsymbol r}}
\def\z{{\boldsymbol z}}

\def\v{{\boldsymbol v}}

\def\Q{{\boldsymbol Q}}

\def\boldeta{\boldsymbol \eta}

\def\bra#1{\langle#1\vert}
\def\ket#1{\vert#1\rangle}

\newcommand{\avg}[1]{\langle #1 \rangle}

\newcommand{\mcal}{\mathcal}

\newcommand{\F}{\mcal{F}}

\def\nab{{\boldsymbol \nabla}}
\def\balpha{{\boldsymbol \alpha}}

\newcommand{\lbr }{\left( }
\newcommand{\rbr }{\right) }
\newcommand{\lc }{\left[ }
\newcommand{\rc }{\right] }
\newcommand{\la }{\left\{ }
\newcommand{\ra }{\right\} }
\newcommand{\hrho }{\hat\rho }
\newcommand{\hH }{\hat H }
\newcommand{\hLmu }{\hat L_\mu }
\newcommand{\Lmu }{L_\mu}
\newcommand{\da}{\dagger}

\newcommand{\D}{\mathcal{D}}

\newcommand{\h}{\mathcal{H}}

\newcommand{\pp}{{\mathbf p}}

\newcommand{\Sg}{{\mathbf S}}
\newcommand{\Id}{{\mathbbm{1}}}

\newcommand*{\multint}{\ensuremath{\int \!\!\!\!\:\int}}
\newcommand*\diff{\mathop{}\!\mathrm{d}}
\providecommand*{\deriv}[3][]{\frac{\diff^{#1}#2}{\diff #3^{#1}}}
\providecommand*{\pderiv}[3][]{\frac{\partial^{#1}#2}{\partial #3^{#1}}}
\providecommand*{\eu}{\ensuremath{\mathrm{e}}}
\providecommand*{\iu}{\ensuremath{\mathrm{i}}}

\begin{document}

\title{Fate of in-medium heavy quarks via a Lindblad equation}

\author[1]{Davide De Boni}

\affiliation[1]{Department of Physics, College of Science, Swansea University, \\ Swansea SA2 8PP, United Kingdom}


\abstract{
What is the dynamics of heavy quarks and antiquarks in a quark gluon plasma? Can heavy-quark bound states dissociate? Can they (re)combine?
These questions are addressed by investigating a Lindblad equation that describes the quantum dynamics of the heavy quarks in a medium.
The Lindblad equations for a heavy quark and a heavy quark-antiquark pair are derived from the gauge theory, following a chain of well-defined approximations. In this work the case of an abelian plasma has been considered, but the extension to the non-abelian case is feasible.
A one-dimensional simulation of the Lindblad equation is performed to extract information about bound-state dissociation, recombination and quantum decoherence for a heavy quark-antiquark pair. All these phenomena are found to depend strongly on the imaginary part of the inter-quark potential.
}

\emailAdd{d.de-boni.840671@swansea.ac.uk, dade89.10@gmail.com}

\date{\today}


\arxivnumber{1705.03567}

\maketitle

\section{Introduction}
\label{sec:intro}

Experimental and theoretical physicists have long sought to understand the properties of the elementary particles (quarks and gluons) described by QCD at extreme temperatures and densities. Experimentalists are creating these extreme conditions by means of relativistic heavy-ion collisions performed at LHC and RHIC.
During these collisions, a fireball of deconfined quarks and gluons, known as the quark gluon plasma (QGP), is formed for a very brief time, which makes the investigation of the fireball a challenging task.
Joint experimental and theoretical efforts have been carried out for about fourty years \cite{Shuryak:1978ij} to find and analyse accessible observables
for diagnosing this short-lived QGP.\\
One of the promising candidates for diagnostics are heavy quarks produced and partecipating in the collisions. Due to the separation of scales between the heavy quarks and the main plasma constituents, the former can be used as dynamical probes of the surrounding medium.
In particular, the seminal work of Matsui and Satz \cite{Matsui:1986dk} suggested that the analysis of bound states of heavy quarks (quarkonia) could lead to a better understanding of the QGP,
the reason being that, at sufficiently high temperatures, colour screening weakens or even prevents a $q\bar q$ pair from binding in the deconfined medium, resulting in a suppression of quarkonia (such as $c\bar c$ and $b\bar b$ bound states) at the  hadronization of the QGP.
This paved the way for many quests for understanding the fate of quarkonia.\\
Some of them made use of screened potential models \cite{Mocsy:2013syh}, based on quantities like the free or internal energies.
The screened potential, whose short- and medium-range parts can be obtained from resummed perturbative calculations, was also used to determine
stationary states of a Schr\"odinger equation \cite{Karsch1988}.
Non-perturbative studies on the lattice followed a different line aiming at reconstructing spectral functions of quarkonia around the crossover temperature \cite{Petreczky:2012rq,Aarts:2011sm,Aarts:2013kaa,Aarts:2014cda}, whereas other approaches exploit the AdS/CFT correspondence \cite{PhysRevD.74.085012,Herzog:2006gh,Albacete:2008dz,Hayata:2012rw}. 
Recent progresses showed that the inter-quark potential not only gets screened at high temperatures, but develops an imaginary part \cite{Laine:2006ns,Burnier:2007qm}, which reflects the Landau damping mechanism due to the collisions between the heavy quarks and plasma constituents \cite{Beraudo:2007ky,Brambilla:2008cx,Brambilla:2010vq}. The fact that the inter-quark potential is actually complex poses serious issues on the validity of the approaches relying on the standard Schr\"odinger equation, which is a good approximation only when the state lives long enough, that is when the deviation from the unitary evolution is small. This happens when the imaginary part of the potential (decay rate) is much smaller than the mass and binding energy of respectively a heavy quark and a heavy-quark bound state\footnote{Moreover the Schr\"odinger equation can not describe possible bound-state recombinations.}. However, the imaginary part of the potential can be used to compute spectral functions of in-medium quarkonia or in the context of the stochastic Schr\"odinger equation (SSE) \cite{PhysRevD.13.857,Gisin1989,0305-4470-26-9-019}.
Both the real and the imaginary part of the potential have been calculated in perturbation theory, but there have also been attempts to get these quantities using lattice \cite{Rothkopf:2011db} and holographic \cite{Noronha:2009da,Giataganas:2013lga,Fadafan:2013coa,0954-3899-43-9-095001} techniques. Some reviews on recent developments on the study of heavy quarks in a quark gluon plasma can be found in \cite{Brambilla:2010cs,Rapp:2009my,CasalderreySolana:2011us,Aarts:2016hap}.\\
The fact that the inter-quark potential is screened and features an imaginary part fosters the phenomenon of quarkonia suppression, which has indeed been partially detected for $J/\Psi$ (see reviews \cite{Rapp:2008tf,Kluberg:2009wc}) and $\Upsilon$ states \cite{Khachatryan:2016xxp} in heavy-ion collision experiments. On the other hand, evidence of recombination of bound states has also been observed for $J/\Psi$ \cite{Abelev:2013ila} (the phenomenon is much less important for $\Upsilon$ \cite{CMS:2017ucd}), hindering the opposite mechanism of dissociation. In fact it is not surprising that some of the heavy particles liberated from the bound states can recombine to form a bound state before the freeze-out of the QGP, especially when there are many heavy quarks and antiquarks tossing about in the hot medium.\\
However, in order to describe the recombination process, a dynamical treatment of the heavy particles becomes necessary, and a static description of the heavy quarks, either via lattice techniques in imaginary time or potential models, is not appropriate anymore. Formulating it as a real-time problem of probes immersed in an environment, the correct language to use is then the one of \textit{open quantum systems}.\\
The goal of this manuscript is to provide a unifying framework to study both dissociation and recombination mechanisms within the language of open quantum systems.
The main idea is to exploit the weak-coupling expansion and the separation of scales between heavy probes and environment to construct an effective theory from the underlying gauge theory, after integrating out the plasma degrees of freedom. A master equation of the Lindblad type for the density matrix of the heavy particles is derived, in which all the quantities appearing are obtained from the complex potential. This equation allows one to study numerically the real-time dynamics of the heavy quarks and antiquarks.\\
Similar strategies based on a Lindblad equation have been used before using a different formalism with \cite{Brambilla:2016wgg} and without \cite{Akamatsu:2012vt,Akamatsu:2014qsa} numerical computations. A one-dimensional numerical simulation has been performed for the master equation and the related SSE in \cite{Akamatsu:2011se}, but the master equation was not directly obtained from the gauge theory. 
The SSE for in-medium quarkonium has been investigated in \cite{Akamatsu:2011se,Rothkopf:2013ria,Rothkopf:2013kya,Kajimoto:2017rel}, but there seems to be no clear way to incorporate dissipative effects (equivalent to the classical friction) within this formalism.
The Schr\"odinger-Langevin equation (SLE) \cite{Katz:2015qja,Katz:2015edb} includes dissipation due to friction but it is not associated to a master equation (hence to a SSE). This means that it is not clear how to derive a SLE from the underlying theory.
Another strategy for studying the dynamics of in-medium quarkonium exploited a classical Fokker-Planck and Langevin equation \cite{PhysRevC.79.034907,Young:2009tj,Moore:2004tg,Akamatsu:2015kaa}.  A generalised Langevin equation for many heavy particles, obtained from the underlying gauge theory, has been simulated in \cite{Blaizot:2015hya}. The limitation of the Langevin equation is that it describes the classical dynamics of the heavy particles, but cannot address topics like quantum decoherence and quantum bound states.\\
This paper extends the semiclassical results of the previous work \cite{Blaizot:2015hya} to a quantum level, by deriving a quantum master equation in the
Lindblad form, which ensures positivity of the density operator, for the density matrix of heavy quarks and antiquarks. 
This equation allows one to follow the quantum dynamics of the heavy particles and compute survival and formation probabilities of heavy-quark bound states, as well as quantum decoherence.\\
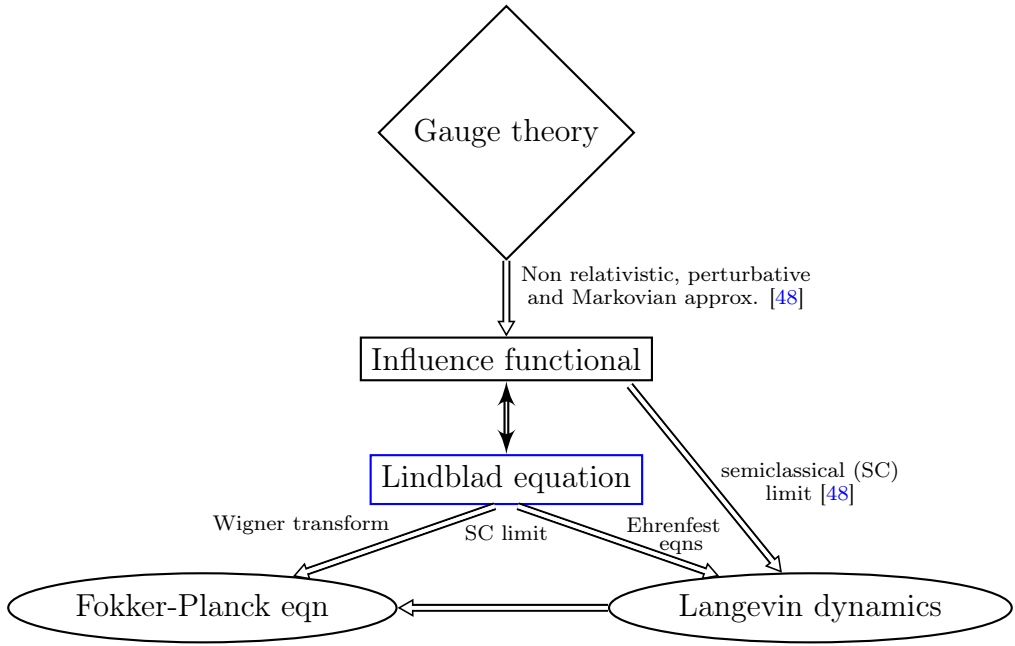
\begin{figure}[t!]
\begin{center}
\begin{tikzpicture}[thick]
  \node at (0,1) [diamond,draw] (g) {Gauge theory};
  \node at (2.1,-0.9) [] (nr) {\scriptsize Non relativistic, perturbative};
  \node at (2.1,-1.2) [] (m) {\scriptsize and Markovian approx. \cite{Blaizot:2015hya}};
  \node at (0,-2) [rectangle,draw] (p) {Influence functional};
  \node at (1.5,-2.2) [] (p2) {}; 
  \node at (4,-3.5) [] (sc) {\scriptsize semiclassical (SC)};
  \node at (4,-3.8) [] (sc) {\scriptsize limit \cite{Blaizot:2015hya}};
  \node at (4,-5.3) [ellipse,draw] (l) {Langevin dynamics};
  \node at (-4,-5.3) [ellipse,draw] (fp) {Fokker-Planck eqn};
  \node at (0,-3.6) [rectangle,draw,blue] (Lin) {{\color{black}Lindblad equation}};
  \node at (0,-3.9) [] (Lin2) {}; 
  \node at (0,-4.3) [] (sc) {\scriptsize SC limit};
  \node at (-2.7,-4.2) [] (sc) {\scriptsize Wigner transform};
  \node at (2.2,-4.2) [] (sc) {\scriptsize Ehrenfest};
  \node at (2.3,-4.5) [] (sc) {\scriptsize eqns};

  \draw[vecArrow] (g) to (p);
  \draw[vecArrow] (p2) to (l);
  \draw[latex'-latex',double] (p) to (Lin);
  \draw[vecArrow] (Lin2) to (l);
  \draw[vecArrow] (Lin2) to (fp);
  \draw[vecArrow] (l) -- (fp);

\end{tikzpicture}
\end{center}
\caption{Flow chart of both the steps performed in the previous work \cite{Blaizot:2015hya} and the ones followed in this project. In general, an influence functional can be derived without the approximations indicated here.}
\label{fig:diagram}
\end{figure}
The flow chart in Fig.\ref{fig:diagram} serves as a guide to understanding how this paper is organised.
Mimicking what has been done in \cite{0305-4470-30-11-030}, section \ref{sec:settings} shows how to build a propagator (or conditional probability), which is related to the influence functional, starting from a generic Lindblad equation.
Section \ref{sec:1Qinf} introduces the model used here to study heavy particles in a thermal bath of light particles. The heavy particles are regarded as being nonrelativistic, whereas the light particles of the medium are treated relativistically. As long as only unbound heavy quarks are concerned, this means that their momenta have to be much smaller than their masses. Similarly, for quarkonia the nonrelativistic condition entails that their relative velocities are much smaller than the speed of light. This paper ignores the specifics of QCD interactions (like attractive/repulsive interactions according to the singlet/octet state for example) and retains only Coulomb interactions. Within this framework, the influence functional for a single heavy quark is obtained with the help of some well-defined approximations, such as the weak coupling and small frequency (Markovian limit) expansions.
In section \ref{sec:1Q} the associated Lindblad equation is then read off from the influence functional.
The Lindbladian terms will result in an expression written solely in terms of the imaginary part of the potential. From the structure of the Lindblad equation, it will be shown that different types of dynamics of the heavy particle are obtained according to the ratio between the correlation length of the environment and the typical size of the heavy probe.
In particular, it will be proven that the classical Langevin dynamics stems from a semiclassical approximation of the Lindblad equation (see lower part of Fig.\ref{fig:diagram}). Section \ref{sec:2Q} contains the computation of the Lindblad equation for a heavy quark-antiquark pair, which could in principle be generalised to $N$ heavy quarks and antiquarks.
Section \ref{sec:modelsim} shows how quantum decoherence and the probability of having a $q\bar q$ bound state at a certain time can be calculated from the density matrix. This section also contains the setup for the one-dimensional numerical simulation of the Lindblad equation for the relative motion of the $q\bar q$ pair, after the motion of the centre of mass has been traced out. The results are displayed in section \ref{sec:numerical}, with specific focus on the linear entropy (a measure of decoherence) and bound-state probabilities, which can be used to extract dissociation and recombination probabilities for different correlation lengths of the thermal bath.
Conclusions and perspectives are found in section \ref{sec:end}, followed by appendices \ref{A3}, \ref{A2}, \ref{A1}.


\section{Density matrix and propagator of an open quantum system}\label{sec:settings}

The goal of this work is to study a probe propagating through a medium in thermal equilibrium. The probe is treated as an open quantum system (also called reduced system or subsystem) that can dissipate or gain energy from the plasma environment. The environment and the probe form a closed quantum system.
A master equation for the density matrix describing the probe will be derived in order to study its real-time dynamics.\\
Of particular interest is the probability $P(\psi, t|\psi_0, t_0)$ to find the reduced system in a quantum state $\ket{\psi}$ at time $t$, given that it was in a state $\ket{\psi_0}$ at time $t_0$. This quantity can be written in terms of the density operator of the subsystem, $\hrho(t) \equiv \mbox{Tr}_{\rm env}\left\{\hrho_{\rm tot}(t)\right\}\,$, where $\hrho_{\rm tot}$ is the density operator of the total (closed) system and the trace is taken over the degrees of freedom of the environment. It is assumed that the subsystem has not yet interacted with the medium at time $t_0$, so that the initial total density operator takes the factorized form 
\be\label{eq:initial}
\hrho_{\rm tot}(t_0)=\hrho(t_0)\otimes \hrho_{\rm env}(t_0)\:,
\ee
with $\hrho(t_0)=|\psi_0\rangle\langle \psi_0|$ and $\hrho_{\rm env}(t_0)=\frac{1}{Z_{\rm env}}\rme^{-\beta \hH_{\rm env}}$ is the density operator of the environment, whose Hamiltonian and partition function are respectively $\hH_{\rm env}$ and $Z_{\rm env}=\mbox{Tr}_{\rm env}\exp(-\beta \hH_{\rm env})$. The total density operator at time $t$ is given by
\be\label{eq:tev}
\hrho_{\rm tot}(t)=\rme^{-i\hH_{\rm tot}(t-t_0)}\hrho_{\rm tot}(t_0)\rme^{i\hH_{\rm tot}(t-t_0)}\:,
\ee
with the Hamiltonian
\be\label{eq:Hamilt}
\hH_{\rm tot}=\hH\otimes{\mathbb{I}}_{\rm env}+{\mathbb{I}}\otimes \hH_{\rm env}+\hH_{\rm int}\:,
\ee
where $\hH$ is the Hamiltonian of the isolated subsystem and $\hH_{\rm int}$ accounts for the interaction between the subsystem and the medium.
The desired probability can now be written as
\be\label{eq:P1}
P(\psi, t|\psi_0, t_0) &=&\int\diff{\phi_{\rm env}}\langle \phi_{\rm env},\psi|\hrho_{\rm tot}(t)|\psi,\phi_{\rm env}\rangle = \mbox{Tr}_{\rm env}\langle \psi|\hrho_{\rm tot}(t)|\psi\rangle\nn\\
&=&\langle \psi|\hrho(t)|\psi\rangle= \int\!\diff{q}\!\!\int\!\diff{q'}\psi(q')\psi^*(q)\rho(t,q,q')\:,
\ee
where $\la\phi_{\rm env}\ra$ forms a basis of the Hilbert space of the medium and $\rho(t,q,q')=\bra{q}\hrho(t)\ket{q'}$ is the density matrix of the subsystem. Using (\ref{eq:tev}), this probability can equivalently be expressed as
\be\label{eq:P2}
&&P(\psi, t|\psi_0, t_0) = \mbox{Tr}_{\rm env}\langle \psi|\rme^{-i\hH_{\rm tot}(t-t_0)}\hrho_{\rm tot}(t_0)\rme^{i\hH_{\rm tot}(t-t_0)}|\psi\rangle\nn\\
&\stackrel{(\ref{eq:initial})}{=}&\int\diff{q_0}\int\diff{q_0'}\,\rho(t_0,q_0,q_0')\mbox{Tr}_{\rm env}\la\langle \psi|\rme^{-i\hH_{\rm tot}(t-t_0)}\hrho_{\rm env}(t_0)|q_0\rangle\langle q_0'|\rme^{i\hH_{\rm tot}(t-t_0)}|\psi\rangle\ra\nn\\
&=&\int\diff{q}\int\diff{q'}\psi(q)\psi^*(q')\int\diff{q_0}\int\diff{q_0'}\,\rho(t_0,q_0,q_0')P(q,q',t|q_0,q_0',t_0)\:,
\ee
where the propagator (or conditional probability) has been defined as
\be\label{eq:1stpropagator}
P(q,q',t|q_0,q_0',t_0) =\mbox{Tr}_{\rm env}\la\langle q|\rme^{-i\hH_{\rm tot}(t-t_0)}\hrho_{\rm env}(t_0)|q_0\rangle\langle q_0'|\rme^{i\hH_{\rm tot}(t-t_0)}|q'\rangle\ra\:.
\ee
Equations (\ref{eq:P1}) and (\ref{eq:P2}) imply that the density matrix is related to the propagator through the relation
\be\label{eq:denprop}
\rho(t,q,q') = \int\diff{q_0}\int\diff{q_0'}\,P(q,q',t|q_0,q_0',t_0)\rho(t_0,q_0,q_0')\:.
\ee
In the next paragraph the path integral expression for the propagator, hence for the density matrix, will be derived in the Markovian limit through the well-known Trotter decomposition. In order to obtain the path integral, one needs to know the equations of motion satisfied by the reduced density operator.
The density operator of the closed system obeys the Liouville-von Neumann equation
\be\label{eq:Liouville}
\iu\hbar\,\frac{d\hrho_{\rm tot}}{dt}(t)=\left[\hH_{\rm tot},\hrho_{\rm tot}(t)\right]\:,
\ee
which is equivalent to a Schr\"odinger equation, describing \textit{reversible} quantum dynamics. Once the trace over the environmental degrees of freedom is taken, the reduced density operator  does not obey the previous equation anymore, being the energy of the open quantum system not conserved.\\
Instead of eq.(\ref{eq:Liouville}), one gets a more complicated equation for $\hrho(t)\,$, known as master equation:
\be\label{eq:DenPro}
\iu\hbar\,\frac{d\hrho}{dt}(t)&=&\mbox{Tr}_{\rm env}\left[\hH_{\rm tot},\hrho_{\rm tot}(t)\right]\stackrel{(\ref{eq:Hamilt})}{=}\left[\hH,\hrho(t)\right]+\mbox{Tr}_{\rm env}\left[{\mathbb{I}}\otimes \hH_{\rm env}+\hH_{\rm int},\hrho_{\rm tot}(t)\right]\nn\\
&\equiv&\left[\hH,\hrho(t)\right]+\iu\,\hat\D\hrho(t)\:,
\ee
where $\hat\D$ is an operator (acting on the Hilbert space of the subsystem) that describes the irreversible part of the dynamics of the subsystem propagating through the medium.


\subsection{Path integral solution of the Lindblad equation}

Here a path integral expression is derived for the reduced density matrix in the Markovian limit, that is when memory effects of the medium can be largely neglected due to the fact that the intrinsic time scale of the subsystem is much larger than the typical bath correlation time. The path-integral construction mimics the one found in \cite{0305-4470-30-11-030}. The same derivation is presented here for self-containment and to set up the language used throughout this work.\\
It has been shown \cite{CIS-338355,8ab9645a29cc4e23a4837264fa594328} that the most general Markovian master equation can be written in the Lindblad form (also known as Bloch equation)
\be
\dot{\hrho} = -\frac{\iu}{\hbar}[\hH,\hrho]+\frac{1}{2\hbar}\sum_\mu\lbr[\hLmu\hrho,\hLmu^\da]+[\hLmu,\hrho\hLmu^\da]\rbr\:,
\ee
or, equivalently,
\be\label{eq:Lindblad}
\dot{\hrho} = \frac{\iu}{\hbar}\lbr-\hH_{\rm eff}\hrho + \hrho\hH_{\rm eff}^\da -\iu\sum_\mu\hLmu\hrho\hLmu^\da\rbr\:,
\ee
with $\hH_{\rm eff}\equiv\hH-\frac{\iu}{2}\sum_\mu\hLmu^\da\hLmu$ and $\la\hLmu\ra_\mu$ forms an arbitrary linear basis of operators acting on the Hilbert space of the subsystem. Starting from the discretised version of the last expression, one finds a relation between
$\hrho(t+\Delta t)$ and $\hrho(t)$ in the position basis, that is
\be\label{eq:timestep}
&&\bra{q}\hrho(t+\Delta t)\ket{q'}= \int_{q_0}\!\int_{q_0'}\bra{q_0}\hrho(t)\ket{q_0'}\lbr \avg{q|q_0}\avg{q_0'|q'}+ \frac{\Delta t}{\hbar}\lbr\vphantom{\sum_\mu}\!\!-\iu\bra{q}\hH_{\rm eff}\ket{q_0}\avg{q_0'|q'}\right.\right.\nn\\
&&\left.\left.+\iu\avg{q|q_0}\avg{q_0'|\hH_{\rm eff}^\da|q'} + \sum_{\mu}\avg{q|\hLmu|q_0}\avg{q_0'|\hLmu^\da|q'} \rbr\rbr
\ee
where the completeness relationship $\int_{q}\ket{q}\bra{q}=\Id$ has been inserted, together with the notation $\int_{q}\equiv\int\diff{q}$. Using that
\be
\avg{q|q_0}=\int_{p}\,\eu^{\frac{i}{\hbar}p(q-q_0)}\;,\qquad\int_{p}\equiv\frac{1}{(2\pi\hbar)^d}\int\!\!\diff{p}\:,
\ee
and the identity
\be\label{eq:Wigner-1}
O(q,q')=\int_{p}\,O\lbr\frac{q+q'}{2},p\rbr\eu^{\frac{i}{\hbar}p(q-q')}\:,
\ee
where the Wigner transform is defined as
\be\label{eq:Wigner}
O(q,p)\equiv \int_{q'}\avg{q-\frac{q'}{2}|\hat O|q+\frac{q'}{2}}\,\eu^{\frac{i}{\hbar}pq'}\:.
\ee
The right hand side of eq.(\ref{eq:timestep}) then reads
\be
&&\int_{q_0}\int_{q_0'}\int_{p}\int_{p'}\bra{q_0}\hrho(t)\ket{q_0'}\,\eu^{\frac{\iu}{\hbar}\lbr p(q-q_0)-p'(q'-q_0')\rbr}\lc 1 + \frac{\iu\Delta t}{\hbar}\cdot\right.\nn\\ 
&&\left.\cdot\lbr -H_{\rm eff}\lbr \bar q,p \rbr + H_{\rm eff}^*\lbr \bar q',p' \rbr 
- \iu\sum_\mu \Lmu\lbr \bar q,p \rbr\Lmu^*\lbr \bar q',p' \rbr\!\rbr\!\!\rc,\nn
\ee
with the midpoints $\bar q\equiv \frac{q+q_0}{2}$ and $\bar q'\equiv \frac{q'+q_0'}{2}$.
Therefore, up to first order in $\Delta t\,$, eq.(\ref{eq:timestep}) re-exponentiated becomes
\be\label{eq:iteration}
&&\bra{q}\hrho(t+\Delta t)\ket{q'}=\int_{q_0}\int_{q_0'}\int_{p}\int_{p'}\bra{q_0}\hrho(t)\ket{q_0'}\exp\lc \frac{\iu}{\hbar}p(q-q_0)-\frac{\iu\Delta t}{\hbar}H_{\rm eff}\lbr \bar q,p\rbr\right.\nn\\
&&\left.-\frac{\iu}{\hbar} p'(q'-q_0')- \frac{\iu\Delta t}{\hbar}H_{\rm eff}^*\lbr \bar q',p'\rbr +\frac{\Delta t}{\hbar}\sum_\mu \Lmu\lbr \bar q,p \rbr\Lmu^*\lbr \bar q',p' \rbr\rc\:.
\ee
The next step consists in deriving the final density matrix $\rho(t,q,q')$ from the initial density matrix $\rho(t_0,q,q')\,$. To accomplish this, the time interval $[t_0,t]$ is partitioned in $N$ small steps of size $\Delta t\,$. Then, after using eq.(\ref{eq:iteration}) iteratively, one is left with
\be
\rho(t,q,q')=\int_{q_0}\int_{q_0'}P(q,q',t|q_0,q_0',t_0)\rho(t_0,q_0,q_0')\:,
\ee 
with the propagator
\be\label{eq:propagator}
&&P(q,q',t|q_0,q_0',t_0)=\lim_{N\to\infty}\int_{p_1}\int_{q_1}\dots\int_{q_{N-1}}\int_{p_N}
\int_{p_1'}\int_{q_1'}\dots\int_{q_{N-1}'}\int_{p_N'}\nn\\
&&\exp\lc \frac{\iu\Delta t}{\hbar}\sum_{i=1}^N\lbr p_i\dot q_i-H_{\rm eff}(\bar q_i,p_i)
-p_i'\dot q_i'+ H_{\rm eff}^*(\bar q_i',p_i')- \iu\sum_\mu \Lmu(\bar q_i,p_i)\Lmu^*(\bar q_i',p_i')\rbr \rc\:,\nn
\ee
and $(q_N,p_N)=(q,p)\,$, $(q_N',p_N')=(q',p')\,$. One should notice that the appearance of the midpoints 
$\bar q_i = \frac{q_i+q_{i-1}}{2}$
is a result of the use of the Wigner representation (\ref{eq:Wigner}). On the other hand the velocities are defined through the retarded rule $\dot q_i=(q_i-q_{i-1})/\Delta t\,$.\\
In the continuous limit the propagator (\ref{eq:propagator}) can be formally written as
\be\label{eq:contpropagator}
P(q_f,q'_f,t|q_0,q_0',t_0) = \int_{(q_0,t_0)}^{(q_f,t)}\!\!\!\D q\D p\!\!\int_{(q_0',t_0)}^{(q'_f,t)}\!\!\!\D q'\D p'\,\exp\lc \frac{\iu}{\hbar}\,S[q,p;q',p'] \rc\:,
\ee
with the phase space action functional
\be\label{eq:genaction}
S[q,p;q',p'] = &&\int_{t_0}^t\diff{\tau}\lc\vphantom{\sum_\mu} \dot qp-H_{\rm eff}(q,p)-\dot q'p' + H_{\rm eff}^*(q',p')\right.\nn\\
&&\left.-\iu\sum_\mu \Lmu(q,p)\Lmu^*(q',p')\rc\:.
\ee
The appearance of the primed and unprimed coordinates can be traced back to the way the density operator evolves with time (see eq.(\ref{eq:tev})), that is with a time evolution from $t_0$ to $t$ and one in the opposite direction. The unprimed coordinates live in the former interval, whereas the primed ones live in the latter branch. These two oriented time branches form the well-known Keldysh contour. Observe that the primed and unprimed coordinates decouple in the absence of Lindblad operators.


\subsection{Equations of motion and quantum decoherence}\label{sec:qd}

When the semiclassical limit $\hbar\to 0$ of expression (\ref{eq:genaction}) is taken, the variational principle
\be
\delta S[q,p;q',p'] = 0
\ee
gives the complex equation of motions corresponding to the stationary paths of the subsystem, that is,
\be\label{eq:eqmotion}
&&\dot q = \pderiv{H_{\rm eff}(q,p)}{p} + \iu\sum_\mu\pderiv{\Lmu(q,p)}{p}\Lmu^*(q',p')\:,\nn\\
&&\dot p = -\pderiv{H_{\rm eff}(q,p)}{q} - \iu\sum_\mu\pderiv{\Lmu(q,p)}{q}\Lmu^*(q',p')\:,\nn\\
&&\dot q' = \pderiv{H_{\rm eff}^*(q',p')}{p'} - \iu\sum_\mu\Lmu(q,p)\pderiv{\Lmu^*(q',p')}{p'}\:,\nn\\
&&\dot p' = -\pderiv{H_{\rm eff}^*(q',p')}{q'} + \iu\sum_\mu\Lmu(q,p)\pderiv{\Lmu^*(q',p')}{q'}\:,
\ee
with the boundary conditions
\be
q(t_0)=q_0\:,\quad q'(t_0)=q_0'\:,\quad q(t)=q_f\:,\quad q'(t)=q_f'\:,
\ee
The equations of motion for one heavy quark propagating through a medium will be derived in section \ref{sec:singleparticle}. Notice the remarkable fact that dissipative dynamics can be derived from a variational principle.
Following \cite{0305-4470-30-11-030}, the propagator (\ref{eq:contpropagator}) can be rewritten in a more transparent way, which separates the reversible dynamics generated by $H(q,p)$ from the irreversible contribution coming from the Lindbladian matrix elements $\Lmu(q,p)\,$. Denoting the Wigner transform of $\hLmu^\da\hLmu$ as $|\Lambda_\mu(q,p)|^2\,$, which implies that
\be
H_{\rm eff}(q,p)=H(q,p)-\frac{\iu}{2}\sum_\mu|\Lambda_\mu(q,p)|^2\:,
\ee
the propagator can be expressed in terms of the classical action of the isolated subsystem,
\be
S_{\rm cl}[q,p]= \int_{t_0}^t\diff{\tau}\lbr \dot qp-H(q,p)\rbr\:,
\ee
and the in-medium piece
\be\label{eq:F}
\F[q,p;q',p'] = \exp\lc \frac{\iu}{\hbar}\varphi[q,p;q',p'] -\frac{1}{2\hbar}D^2[q,p;q',p'] \rc\:,
\ee 
where
\be\label{eq:varphi}
\varphi[q,p;q',p'] = \sum_\mu\int_{t_0}^t\!\!\!\diff{\tau}\,\mbox{Im}\lc \Lmu(q,p)\Lmu^*(q',p') \rc\:,
\ee
is a phase functional modifying the classical action of the isolated system and
\be\label{eq:D}
D[q,p;q',p'] = &&\lbr \sum_\mu\int_{t_0}^t\diff{\tau} \lc\frac{}{} |\Lambda_\mu(q,p)|^2 + |\Lambda_\mu(q',p')|^2 \right.\right.\nn\\
&&\left.\left.- 2\mbox{Re}\lc \Lmu(q,p)\Lmu^*(q',p') \rc \frac{}{}\rc \vphantom{\sum_\mu}\rbr^{1/2}
\ee
is called decoherence distance functional, since it measures the decoherence between the phase space histories \cite{0305-4470-30-11-030}.
Notice that the phase functional $\varphi$ has to vanish when one takes $(q',p')=(q,p)\,$, and $D$ is symmetric under $(q',p')\leftrightarrow(q,p)$ exchange.\\
The propagator (\ref{eq:contpropagator}) can be finally written in terms of the functionals $\varphi$ and $D$ as
\be\label{eq:Pphasespace}
&&P(q_f,q'_f,t|q_0,q_0',t_0) \nn\\
&&=\int_{(q_0,t_0)}^{(q_f,t)}\!\!\!\!\!\!\!\!\!\D q\D p\!\!\int_{(q_0',t_0)}^{(q'_f,t)}\!\!\!\!\!\!\!\!\!\D q'\D p'\exp\lc \frac{\iu}{\hbar}\lbr S_{\rm cl}[q,p] -S_{\rm cl}[q',p']\rbr \rc \F[q,p;q',p'].
\ee
In the following sections the phase functional $\varphi$ and the decoherence functional $D$ will be calculated first for one heavy quark and then for a heavy $q\bar q$ pair propagating through a quark gluon plasma.


\section{Heavy quarks in a thermal medium: the model}\label{sec:1Qinf}

The general setup used in this paper is the same discussed in the previous work \cite{Blaizot:2015hya}. Only the key features of the model (such as physical scales and approximations) are presented here, whereas all the other details and calculations are found in \cite{Blaizot:2015hya}.\\
In order to describe the real-time dynamics of heavy quarks and antiquarks propagating through a plasma of thermalised light particles, an effective theory for the heavy probes is obtained by tracing out the plasma degrees of freedom.\\
The model considered here consists of an abelian plasma of massless charged particles interacting with positive- and negative-charged heavy particles, which are called heavy quarks and antiquarks respectively. Strictly speaking, this picture applies only to electromagnetic interactions, being non abelian features of QCD not taken into account. However, the language of QCD has been used throughout this manuscript due to the similarities between the QED and QCD plasmas at very high temperatures.
In this model $N$-nonrelativistic\footnote{Consequently their number is fixed.} heavy quarks and antiquarks have been considered, whose mass $m$ is much larger than the temperature $T=1/\beta$ of the plasma ($k_{B}$ and $c$ have been set to $1$). The heavy particles are described using first quantisation, whereas the light particles of the thermal bath are considered within a quantum field theory framework.
The heavy probes interact amongst themselves and with the plasma particles via Coulomb interactions. Magnetic interactions among the heavy probes are negligible in the non relativistic limit, being suppressed by powers of the velocity, or $p/m\,$. Magnetic interactions are also not taken into account for the light particles of the medium since the heavy quarks do not induce magnetic excitations.
Therefore the whole system undergoes only Coulomb interactions. In the Coulomb gauge, the Hamiltonian of the total system reads
\be\label{eq:hamiltonian}
H_{\rm tot} &=& \frac{1}{2m}\sum_{j=1}^N\left(\pp_j^2 + \bar\pp_j^2\right) + \int\diff{\x}~{\psi}^\dagger(\x)~\left(\frac{\hbar\,\balpha\cdot \nab}{i}+m_\psi\gamma_0\right)~\psi(\x) +\nn\\
&+& \frac{\hbar}{2}\multint\diff{\x}\diff{\y}~j_{\rm tot}^0(\x)\frac{1}{4\pi|\x-\y|}j_{\rm tot}^0(\y),
\ee
where $\alpha^i=\gamma_0 \gamma^i$ is a Dirac matrix, and  $j_{\rm tot}^0=j^0+j_\psi^0$ is the total charge density, with 
\be\label{rho}
j^0(\x) = g\,\sum_{j=1}^N\left[ \delta(\x- \q_j) -\delta(\x-\bar\q_j)\right]\:,
\ee 
the charge density of the heavy quarks and antiquarks, and
\be
j_\psi^0(\x)=g\,\psi^{\dagger}(\x)\psi(\x)
\ee
the density of the charged light quarks and antiquarks of the plasma, where $g$ is the gauge coupling.
The spatial coordinates $\q_j$ and $\bar\q_j$, with $j=1,\cdots,N$, refer to the heavy quarks and antiquarks respectively, and $\p_j$, $\bar\p_j$ are the corresponding momenta.
The plasma is supposed to be electrically neutral, that is, it contains the same number of light quarks and antiquarks.


\subsection{Influence functional for $N$ heavy quarks and antiquarks}

In \cite{Blaizot:2015hya} it has been shown in details how to get the path integral expression of the propagator (\ref{eq:1stpropagator}) for the heavy particles, after integrating out the plasma degrees of freedom. The conditional probability can be written as
\be\label{pathint}
P(\Q_f,\Q'_f, t|\Q_0,\Q_0', t_0)=\int_{(\Q_0,t_0)}^{(\Q_f,t)}\!\!\!\!\!\!{\cal D}\Q\int_{(\Q_0',t_0)}^{(\Q'_f,t)}\!\!\!\!\!\!{\cal D}\Q'\,\eu^{\frac{\iu}{\hbar}\,\left(S_0[\Q,\Q']+g^2\Phi[\Q,\Q']+\,\mbox{\scriptsize o}(g^4)\right)}\:,
\ee
where the $2N$-dimensional vector $\Q=\left(\q_1,\cdots,\q_N,\bar\q_1,\cdots,\bar\q_N\right)$ (and similarly for $\Q'$) collectively denotes the spatial coordinates of all the heavy particles.
The coordinate $\Q$ lives in the first (upper) branch of the Keldysh contour in the complex time plane. This branch runs from $\tau=0$ to $\tau=t\,$. The coordinate $\Q'$ lives in the second (lower) branch, which runs backwards in time from $\tau=t$ to $\tau=0\,$.
The free action of the heavy particles is
\be
S_0[\Q,\Q']=\frac{m}{2}\sum_{i=1}^N\int_{t_0}^{t}\diff{\tau}\left(\dot{\q}_{i}^2-\dot{\q}_{i}^{'2}+ \dot{\bar\q}_{i}^2-\dot{\bar\q}_{i}^{'2}\right)\:.
\ee
The influence functional $\Phi$ is obtained after performing a perturbative expansion up to order $\alpha_s=g^2/(4\pi)$ (non-linear interactions are ignored) and considering almost instantaneous interactions between heavy and light particles, exploiting the fact that one is interested in studying the dynamics of the heavy probes over a time scale much larger than the typical one of the particles of the medium\footnote{Refer to the small frequency expansion of the correlation functions in \cite{Blaizot:2015hya} for further details and the derivation of the influence functional.}.
The influence functional can be split into three terms as  $\Phi = \Phi_{_{QQ}}+\Phi_{_{\bar Q\bar Q}}+\Phi_{_{Q \bar Q}}\,$, where
\be\label{QQ}
\Phi_{_{QQ}}[\Q,\Q'] &=&
\frac{1}{2} \sum_{i,j=1}^N \int_{t_0}^{t} \diff{\tau}
\left[\frac{\!}{\!}V(\q_{j}-\q_{i})-V(\q_{j}'-\q_{i}')\right.\nn\\
&&-\iu W(\q_{j}'-\q_{i}')-\iu W(\q_{j}-\q_{i})+2\,\iu\,W(\q_{j}-\q_{i}')\nn\\
&&+\left. \frac{\beta\hbar}{2}(\dot{\q}_{j}+\dot{\q}_{i}')\cdot\frac{\partial}{\partial \q_{i}'} W(\q_{j}-\q_{i}')\right].
\ee
and similarly for $\Phi_{_{\bar Q\bar Q}}$ with the substitution $\{\q_i\}\rightarrow\{\bar\q_i\}$. The mixed quark-antiquark contribution reads
\be\label{eq:QbarQ}
\!\!\!\!\!\!\!\!\!\!\!\!\!\!\!\!\!\!\!\!\!&&\Phi_{_{_{Q \overline Q}}}[\Q,\Q'] =-\sum_{i,j=1}^N \int_{t_0}^{t} \diff{\tau}
\left[\frac{\!}{\!} V(\q_{j}-\bar\q_{i})-V(\q_{j}'-\bar\q_{i}')\right.\nn\\
&&-\iu W(\q_{j}'-\bar\q_{i}')-\iu W(\q_{j}-\bar\q_{i})+\iu W(\q_{j}-\overline\q_{i}')+\iu W(\q_{j}'-\bar\q_{i})\\
&&+\left. \frac{\beta\hbar}{4}\left((\dot{\bar\q}_{i}'+\dot\q_j)\cdot\pderiv{W(\q_{j}-\bar\q_{i}')}{\bar\q_{i}'}- (\dot{\bar\q}_{i}+\dot\q_j')\cdot\pderiv{W(\bar\q_{i}-\q_{j}')}{\bar\q_{i}}\right)\right]\:.\nn
\ee
The functions $V$ and $W$ come from the correlation functions describing the interactions between the heavy probes amongst themselves and with the bath. They are respectively the real and the imaginary part of the complex potential experienced by the heavy particles, derived as
\be\label{W}
V(\x)&=&-\frac{\hbar}{4\pi\,x}\,\eu^{-\frac{m_{_{\scriptsize D}}}{\hbar}x}\:,\nn\\
\nn\\
W(\x)&=&-\frac{\hbar T}{2\pi\,m_{_{\scriptsize D}}x}\int_0^{\frac{\Lambda}{m_{_{\scriptsize D}}}}\!\!\!\diff{z}\,\frac{\sin\lc z\frac{m_{_{\scriptsize D}}x}{\hbar}\rc}{(z^2+1)^2}\:,
\ee
where $x=|\x|\,$, $m_{_{\scriptsize D}}\propto gT$ is the Debye mass and $\Lambda$ is a momentum cutoff coming from the Hard Thermal Loop approximation (see discussion in \cite{Blaizot:2015hya}). 
In the next paragraph, all the terms appearing in the influence functional will be related to the Lindblad matrix elements introduced in section \ref{sec:qd}.
This is the point where this project takes another path compared to the one followed in the previous work \cite{Blaizot:2015hya}.
In the latter, a semiclassical approximation of $\Phi$ was performed in order to get a generalised Langevin equation for $N$ heavy quarks and antiquarks.
Instead, here a quantum master equation will be derived from the influence functional (see Fig.\ref{fig:diagram}). The master equation allows one to deal with quantum objects like bound states and study related quantum phenomena. These aspects could not be studied with a classical Langevin equation.


\subsection{Conditional probability for a single heavy quark}\label{sec:singleparticle}

For a single heavy quark in the plasma the influence functional simplifies to
\be\label{eq:singleQ}
\Phi[\q,\q'] =
\int_{t_0}^{t} \diff{\tau}\left[\iu\lbr W(\q-\q')-W(0)\rbr +\frac{\beta\hbar}{4}\frac{\partial}{\partial \q'} W(\q-\q')\cdot(\dot{\q}+\dot{\q}')\right].
\ee
In order to calculate the phase functional $\varphi$ and the decoherence distance functional $D$ for a single heavy quark, the conditional probability (\ref{pathint}) has to be rewritten in phase space as in (\ref{eq:Pphasespace}). To achieve this, the following identity is useful:
\be\label{eq:gaussidentity}
&&\int\diff{p}\int\diff{p'} \exp\lc \frac{\iu}{\hbar}\lbr \frac{p'^2}{2m}-\frac{p^2}{2m}+pf(q,q',\dot q)-p'h(q,q',\dot q') \rbr \rc \nn\\
&=& (2\pi m\hbar) \exp\lc \frac{\iu m}{2\hbar}\lbr f^2(q,q',\dot q)-h^2(q,q',\dot q') \rbr\rc\:.
\ee
Now one defines the functions
\be
&&f(q,q',\dot q)=\dot q +\frac{g^2\beta\hbar}{4m}\frac{\partial}{\partial q'} W(q-q')\nn\\
&&h(q,q',\dot q')=\dot q' - \frac{g^2\beta\hbar}{4m}\frac{\partial}{\partial q'} W(q-q')\:,
\ee
and neglect terms of order $g^4$ when computing $f^2$ and $h^2\,$. It is then easy to prove that the propagator can be expressed in phase space as in (\ref{eq:Pphasespace}), where the classical action is
\be
S_{\rm cl}[\q,\p]= \int_{t_0}^t\diff{\tau}\lbr \dot \q\p-\frac{\p^2}{2m}\rbr\:,
\ee
and the in-medium contribution
\be\label{eq:Fviol}
&&\!\!\!\!\!\!\!\!\!\!\F[\q,\p;\q',\p']=\\
&&\!\!\!\!\!\!\!\!\!\!=\exp\lc \frac{\iu}{\hbar}g^2\!\!\int_{t_0}^{t}\!\!\diff{\tau}\!\lc \iu(W(\q-\q')-W(0)) + \frac{\beta\hbar}{4m}(\p+\p')\cdot\frac{\partial}{\partial \q'} W(\q-\q') \rc\!\!+\mbox{o}(g^4)\rc\:.\nn
\ee
One can show that the Lindblad equation that generates the propagator (\ref{eq:Pphasespace}) with (\ref{eq:Fviol}) violates the conservation of probability ($\deriv{}{t}\mbox{Tr}\,\hrho(t)=0\,$) to order $g^4$.
Hence, with hindsight, the following term, which is negligible both in perturbation theory and in the non relativistic limit ($T/m\ll 1\,$), is added ad hoc to the in-medium part $\F\,$, in order to exactly preserve probability:
\be\label{eq:negl}
\frac{\iu g^2\beta\hbar^2}{2m}\pderiv[2]{W(\q-\q')}{\q}=\frac{\iu g^2 m_{_{\scriptsize D}}^2}{2mT}\pderiv[2]{W(\z)}{\z}\sim \frac{g^4T}{m}W\:,
\ee
where $\z\equiv\frac{m_{_{\scriptsize D}}}{\hbar}(\q-\q')$ is a dimensionless parameter.
The corrected functional then reads
\be\label{eq:F1Q}
\F[\q,\p;\q',\p']&=&\exp\lc \frac{\iu}{\hbar}g^2\!\!\int_{t_0}^{t}\!\!\diff{\tau}\!\lc \iu(W(\q-\q')-W(0))-\frac{\iu\beta\hbar^2}{2m}\pderiv[2]{W(\q-\q')}{\q} \right.\right.\nn\\
&&\left.\left.+ \frac{\beta\hbar}{4m}(\p+\p')\cdot\frac{\partial}{\partial \q'} W(\q-\q') \rc\!\!+\mbox{o}(g^4)\rc\:.
\ee
Finally eq.(\ref{eq:F1Q}), (\ref{eq:F}), (\ref{eq:varphi}) and (\ref{eq:D}) give the phase and distance decoherence functionals for the case of a single heavy quark in the medium:
\be\label{eq:varphi1Q}
\varphi[\q,\p;\q',\p'] &=& \sum_\mu\int_{t_0}^t\!\!\!\diff{\tau}\,\mbox{Im}\lc \Lmu(\q,\p)\Lmu^*(\q',\p') \rc\nn\\
&=& \frac{g^2\beta\hbar}{4m}\int_{t_0}^{t}\!\!\diff{\tau}\:(\p+\p')\cdot\frac{\partial}{\partial \q'} W(\q-\q')\:,
\ee
and
\be\label{eq:D1Q}
D^2[\q,\p;\q',\p'] &=& \sum_\mu\int_{t_0}^t\diff{\tau} \lc\vphantom{\sum} |\Lambda_\mu(\q,\p)|^2 + |\Lambda_\mu(\q',\p')|^2 - 2\mbox{Re}\lc \Lmu(\q,\p)\Lmu^*(\q',\p') \rc \rc\nn\\
&=& 2g^2\int_{t_0}^{t}\!\!\diff{\tau}\lc(W(\q-\q')-W(0))-\frac{\beta\hbar^2}{4m}\pderiv[2]{W(\q-\q')}{\q}\rc\nn\\
&=&D^2[\q;\q']\:.
\ee
Notice that $\left.\pderiv{W}{\x}\right|_{\x=0}=0$ gives the required property $\varphi[\q,\p;\q,\p]=0\,$. Moreover, the fact that $W$ is an even function makes the decoherence distance satisfy the symmetric property $D^2[\q,\p;\q',\p']=D^2[\q',\p';\q,\p]\,$. Equations (\ref{eq:varphi1Q}) and (\ref{eq:D1Q}) show explicitly how the Lindbladian terms are related to the imaginary part of the potential.
Combining the first lines of (\ref{eq:varphi1Q}) and (\ref{eq:D1Q}), one also obtains the general relation\footnote{The fact that here $D$ does not depend on the momenta can be understood, for example, in the case of a Lindblad operator of the form
$
\hat L = \sigma \hat f(\hat\q) +\iu g^2\hat\p\:,$ with $\sigma,g^2\in\mathbbm{R}\,$, $\hat f^\da=\hat f$ and negligible o$(g^4)$ terms.
This Lindblad operator gives $|\Lambda(\q,\p)|^2=|\Lambda(\q)|^2+\,\mbox{o}(g^4)$ and $\mbox{Re}\lc L(\q,\p)L^*(\q',\p')\rc = \mbox{Re}\lc L(\q)L^*(\q')\rc+\,\mbox{o}(g^4)\,
$. This is why eq.(\ref{eq:LL*}) assumes $\Lambda_\mu$ to be momentum independent.}
\be\label{eq:LL*}
&&\sum_\mu\Lmu(\q,\p)\Lmu^*(\q',\p')=\nn\\
&&\iu\varphi_{\tau}[\q,\p;\q',\p']-\frac{1}{2}D^2_{\tau}[\q;\q'] + \frac{1}{2}\sum_\mu\lbr\vphantom{\sum}|\Lambda_\mu(\q)|^2 + |\Lambda_\mu(\q')|^2\rbr \:,
\ee
where the definitions $\varphi\equiv\int_{t_0}^{t}\!\!\diff{\tau}\varphi_{\tau}\,$, $D^2\equiv\int_{t_0}^{t}\!\!\diff{\tau}D^2_{\tau}$
have been used. Relation (\ref{eq:LL*}) will be useful to derive the Lindblad equation for the density matrix.\\
Moreover, from (\ref{eq:LL*}) and minimising the action, one obtains the classical equations of motion (\ref{eq:eqmotion}) for the single quark case. It is interesting to observe that the Lindbladian terms modify the equations of motion from the standard $m\dot\q=\p$ and $m\dot\q'=\p'$ to
\be
&&\!\!\!\!\!\dot\q = \frac{\p}{m}-\pderiv{\varphi_{\tau}}{\p}=\frac{\p}{m}-\frac{g^2\beta\hbar}{4m}\frac{\partial}{\partial \q'} W(\q-\q')\:,\\
&&\!\!\!\!\!\dot\q' = \frac{\p'}{m}+\pderiv{\varphi_{\tau}}{\p'} = \frac{\p'}{m}+\frac{g^2\beta\hbar}{4m}\frac{\partial}{\partial \q'} W(\q-\q')\:.
\ee
This will turn out to be a crucial point when a Fokker-Planck-like equation for the Wigner function of a $q\bar q$ pair is derived from a Langevin equation (see Appendix \ref{A1}).


\newpage


\section{Lindblad equation for one heavy quark}\label{sec:1Q}

This section shows how to get the Lindblad equation for the density matrix $\rho(t,\q,\q')$ starting from the Lindblad equation for the density operator $\hrho\,$. The first step is to project eq. (\ref{eq:Lindblad}) into position basis: 
\be\label{eq:Lindblad2}
\bra{\q}\dot{\hrho}\ket{\q'} = \frac{\iu}{\hbar}\lbr-\bra{\q}\hH_{\rm eff}\hrho\ket{\q'} + \bra{\q}\hrho\hH_{\rm eff}^\da\ket{\q'} -\iu\sum_\mu\bra{\q}\hLmu\hrho\hLmu^\da\ket{\q'}\rbr\:.
\ee
The intermediate steps of the calculation are shown in Appendix \ref{A3}. Only the main results are reported here. 
For a heavy quark in the plasma (without external potential) the first two terms on the right-hand side read
\be\label{eq:1}
-\bra{\q}\hH_{\rm eff}\hrho\ket{\q'}&=&-\int_{\x}\bra{\q}\hH_{\rm eff}\ket{\x}\rho(t,\x,\q')\nn \\
&=& \lbr\frac{\hbar^2}{2m}\pderiv[2]{}{\q} + \frac{\iu}{2}\sum_\mu\left|\Lambda_\mu\lbr \q\rbr\right|^2 \rbr \rho(t,\q,\q')\:,
\ee
and
\be\label{eq:2}
\bra{\q}\hrho\hH_{\rm eff}^\da\ket{\q'}= \lbr -\frac{\hbar^2}{2m}\pderiv[2]{}{\q'} + \frac{\iu}{2}\sum_\mu\left|\Lambda_\mu\lbr \q'\rbr\right|^2 \rbr \rho(t,\q,\q')\:.
\ee
The last term in (\ref{eq:Lindblad2}) gives
\be\label{eq:3}
&&\!\!\!-\iu\sum_\mu\bra{\q}\hLmu\hrho\hLmu^\da\ket{\q'}=\nn\\
&&\!\!\!\int_{\x}\int_{\y}\,\rho(t,2\x-\q,2\y-\q')\int_\p\int_\l\,\varphi_{\tau}[\x,\p/2;\y,\l/2]\,\eu^{\frac{\iu}{\hbar}\p\cdot(\q-\x)}\eu^{-\frac{\iu}{\hbar}\l\cdot(\q'-\y)}\nn\\
&&\!\!\!+\frac{\iu}{2}\lbr D_{\tau}^2[\q;\q']- \sum_\mu\lc|\Lambda_\mu(\q)|^2 + |\Lambda_\mu(\q')|^2\rc \rbr \rho(t,\q,\q')\:.
\ee
Collecting all the results in (\ref{eq:Lindblad2}), (\ref{eq:1}), (\ref{eq:2}) and (\ref{eq:3}), one gets
\be
&&\!\!\!\!\!\!\!\!\!\!\!\!-\iu\hbar\,\dot\rho(t,\q,\q')=\lbr\frac{\hbar^2}{2m}\lbr\pderiv[2]{}{\q}-\pderiv[2]{}{\q'}\rbr+\frac{\iu}{2} D_{\tau}^2[\q;\q'] \rbr \rho(t,\q,\q')\nn\\
&&\!\!\!\!\!\!\!\!\!\!\!\!+\int_{\x}\int_{\y}\,\rho(t,2\x-\q,2\y-\q')\int_\p\int_\l\,\varphi_{\tau}\lc\x,\frac{\p}{2};\y,\frac{\l}{2}\rc\eu^{\frac{\iu}{\hbar}\p\cdot(\q-\x)}\eu^{-\frac{\iu}{\hbar}\l\cdot(\q'-\y)}\:.
\ee
Using relations (\ref{eq:varphi1Q}) and (\ref{eq:D1Q}) for $\varphi_\tau$ and $D_\tau$ respectively, together with integration by parts, the Lindblad equation for the density matrix finally reads
\be\label{eq:Lindblad1Q}
\pderiv{\rho(t,\q,\q')}{t}&&=\lbr\frac{\iu\hbar}{2m}\lbr\pderiv[2]{}{\q}-\pderiv[2]{}{\q'}\rbr-\frac{g^2}{\hbar}(W(\q-\q')-W(0))\right.\nn\\
&&\left.-\frac{g^2\beta\hbar}{4m}\pderiv{W(\q-\q')}{\q'}\cdot\lbr \frac{\partial}{\partial \q'}-\frac{\partial}{\partial \q} \rbr \rbr \rho(t,\q,\q')\:.
\ee
It is immediate to check that probability is preserved by the Lindblad equation:
\be
\deriv{}{t}\mbox{Tr}\,\hrho(t)= \int_{\q}\int_{\q'}\,\delta(\q-\q')\pderiv{}{t}\rho(t,\q,\q') = 0\:,
\ee
being $\left.\partial_\q W(\q)\right|_{\q=\bold{0}}=0\,$.
Notice that (\ref{eq:Lindblad1Q}) is the same Lindblad equation derived in \cite{Akamatsu:2012vt} (see their eq.$(66)\,$) using a different (variational) approach.
It is important to observe that the Lindbladian terms in (\ref{eq:Lindblad1Q}) have little effect on the diagonal elements of the density matrix. On the other hand, it will be shown that the off-diagonal terms become suppressed at long times. The disappearence of the off-diagonal peaks of the density matrix, which are due to quantum interference, is called quantum decoherence. This phenomenon makes the system undergo a transition from quantum to classical dynamics \cite{Zurek2007}.


\subsection{Resolution of the quantum probe by the environment}

Different types of dynamics of the heavy quark in the bath arise according to the ability of the medium to resolve the subsystem. This ability depends on the correlation length of the environment ($l_{env}$) relevant to the heavy quark-environment system, which corresponds to the typical spatial region in which $W(\x)$ is significantly different from zero.
The environment is expected to be able to resolve, hence affect, the system if and only if its correlation length is of the order of the typical size of the quantum system ($l_{sys}$). This can also be understood from the fact that the typical momentum kicks that the medium gives to the probe are $p_{env}\sim h/l_{env}\,$. This means that a very large correlation length allows the environment to distribute only very small momentum kicks\footnote{A possible exception to this argument has to be considered if there are correlated fields in the background instead of random fluctuations.}, therefore the subsystem is barely perturbed. In the limit of $p_{env}=0\,$, the system continues following its reversible dynamics dictated by the Schr\"odinger equation.\\
When $l_{\rm env}\gtrsim l_{\rm sys}$, it will be shown that the system undergoes a Brownian motion, with the bath providing a white noise disturbing the probe.
In the $\!$ $l_{env}\ll l_{sys}$ regime the magnitude of the momentum kicks that the medium gives to the system is much bigger than the typical energies of the latter, and the dynamics of the system is then largely affected by the presence of the environment. In this scenario all bound states are expected to dissociate quickly. Of particular interest is the case with $l_{env}/l_{sys}\lesssim 1\,$, for which the dynamics of the distinct bound states is affected differently by the medium. In this case the fate of the bound states can be regarded as a hadronic calorimeter.
Below, this qualitative physical description will be made more quantitative.


\subsection{Emergence of different types of dynamics}\label{sec:sc1}

An inspection of the Lindblad equation will bring to the conclusion that there are four distinct types of dynamics that the system can exhibit:
\begin{itemize}
\item  $l_{env}/l_{sys}\gg 1$\\
Reversible dynamics;
\item   if $l_{env}/l_{sys}\gtrsim 1$\\
Open-system dynamics.\\ The dynamics becomes classical (Langevin-like) after a decoherence time;
\item  $l_{env}/l_{sys}\lesssim 1$\\
The bath acts like a bound-state sieve;
\item  $l_{env}/l_{sys}\ll 1$\\
The bath does not discriminate amongst bound states; they all decay at the same rate.
\end{itemize}
In order to prove these statements let us first perform the following change of variables:
\be\label{eq:change}
\r\equiv\frac{1}{2}(\q+\q')\:,\qquad\y\equiv\q-\q' \:,
\ee
where the $\y$ coordinate describes the off-diagonal elements of the density matrix.  
The Lindblad equation (\ref{eq:Lindblad1Q}), with an additional external (real) potential $V_{\rm ext}\,$, then reads
\be\label{eq:1particlerho}
\!\!\pderiv{\rho(t,\r,\y)}{t}=&&\lbr\frac{\iu\hbar}{m}\partial_{\r}\cdot\partial_{\y}-\frac{\iu}{\hbar}(V_{\rm ext}(\r+\y/2)-V_{\rm ext}(\r-\y/2))\right.\nn\\
&&\left.-\frac{g^2}{\hbar}(W(\y)-W(0))-\frac{g^2\hbar\beta}{2m}\partial_{\y} W(\y)\cdot\partial_{\y} \rbr \rho(t,\r,\y)\:.
\ee
The external potential has been introduced in order to study the dynamics of bound states.
Notice again that the diagonal elements $\rho(t,\r,\bm{0})$ are almost unaffected by the Lindbladian terms.\\
It is immediate to see that, if $l_{env}/l_{sys}\gg 1\,$, $W(\x)$ is constant in the region where the density matrix is different from zero. This means that all Lindbladian terms go to zero in the region where the subsystem is localised. Therefore the subsystem behaves like a closed quantum system obeying a standard Schr\"odinger equation.\\
In the scenario where $W$ slightly varies over the region where $\rho$ has support, the former can be Taylor expanded for small $|\x|/l_{env}$. The expansion of $W$ up to second order in its argument is called semiclassical approximation and reads
\be
W(\y)=W(0)+\frac{1}{2}\y\cdot\h(0)\cdot\y +\, \mbox{o}(\y^4)\:,
\ee
where $\h(0)$ is the positive-definite Hessian matrix of $W(\y)$ evaluated at $\y=0$ and $\left.\partial_\y W(\y)\right|_{\y=0}=0$ has been used. Assuming that the semiclassical approximation is good enough also for $V_{\rm ext}\,$, the Lindblad equation reads
\be\label{eq:Caldeira1}
&&\partial_t\rho(t,\r,\y)=\\
&&\lbr\frac{\iu\hbar}{m}\partial_\r\cdot\partial_\y-\frac{\iu}{\hbar}\y\cdot\partial_\r V_{\rm ext}(\r)-\frac{g^2}{2\hbar}\y\cdot\h(0)\cdot\y-\frac{g^2\hbar\beta}{2m}\y\cdot\h(0)\cdot\partial_\y \rbr \rho(t,\r,\y)\nn\:.
\ee
Using the fact that $\h(0)$ is a diagonal matrix, and introducing the friction coefficient $\gamma$ and the momentum diffusion coefficient $\kappa$ respectively as (see \cite{Blaizot:2015hya})
\footnote{It will be shown below that the momentum diffusion coefficient $\kappa$ gives the strength of the momentum kicks that the system receives from the medium. The associated (position) diffusion coefficient is $D=\frac{2T^2}{\kappa}=\frac{T}{m\gamma}$. Observe that $D\sim 1/\lbr g^4T\ln\lbr 1/g^2 \rbr\rbr$, like all transport scales in perturbation theory \cite{Arnold:2000dr}. This result comes from the fact that $\gamma\sim \frac{g^4T^2}{m}\ln(1/g^2)$, where the expression (\ref{W}) for $W$ has been used together with $\Lambda^2/m_{_D}^2\sim 1/g^2\gg 1\,$.}
\be\label{eq:fricdiff}
\h(0)_{ij} = \delta_{ij}\frac{2\gamma m}{g^2\hbar\beta}\:,\qquad \kappa = \frac{2m\gamma}{\beta}=2m\gamma T\:,
\ee
one gets an equation of the Caldeira-Leggett type for the density matrix:
\be\label{eq:Caldeira11}
\partial_t \rho(t,\r,\y)=\lbr\frac{\iu\hbar}{m}\partial_{\r}\cdot\partial_{\y}-\frac{\iu}{\hbar}\y\cdot\partial_\r V_{\rm ext}(\r)-\frac{\kappa}{2\hbar^2}\y^2-\gamma\y\cdot\partial_{\y} \rbr \rho(t,\r,\y)\:.
\ee
Notice that the semiclassical approximation preserves the trace of the density matrix.
The penultimate term in eq.(\ref{eq:Caldeira11}) has very little effect on the diagonal peaks of the density matrix, but it causes the off-diagonal ones ($\rho_{\rm off}$)
to decay like  \cite{Zurek2007}
\be
\partial_t \rho_{\rm off} \approx -\frac{\kappa}{2\hbar^2}\y^2\rho_{\rm off}=-\tau_{_D}^{-1}\rho_{\rm off}\:,
\ee
where 
\be
\tau_{_D}=2\tau_{_R}\lbr\frac{\hbar}{\sqrt{2mT}|\y|}\rbr^2=2\tau_{_R}\lbr\frac{\lambda_{\rm th}}{|\y|}\rbr^2
\ee
is the decoherence time, $\tau_{_R}=\gamma^{-1}$ the relaxation time and $\lambda_{\rm th}$ the thermal de Broglie wavelength. For macroscopic objects, the decoherence time is typically much shorter than the relaxation time.
Taking the Wigner transform (see \cite{2008AmJPh..76..937C} for a good review on Wigner functions and their properties)
\be
\rho(t,\r,\p)=\int\!\!\diff{\y}\,\rho(t,\r,\y)\eu^{-\frac{\iu}{\hbar}\p\cdot\y}
\ee
of eq.(\ref{eq:Caldeira11}), one gets the following Fokker-Planck-like equation for the Wigner function:
\be\label{eq:Fokker1}
\lc \partial_{t} + \frac{\p}{m}\cdot\partial_{\r}-\partial_\r V_{\rm ext}(\r)\cdot\partial_{\p}\rc\rho(t,\r,\p)=\gamma\lc  mT\,\partial_\p^2 + \partial_{\p}\cdot\p\rc\rho(t,\r,\p)\:.
\ee
Recall that the Wigner function can be negative and therefore is not a phase space probability. However, after a short time of the order of the decoherence time, the Wigner function becomes positive and can be regarded as a classical probability distribution. In this scenario it is easy to show (see \cite{schwabl2006statistical}) that the above Fokker-Planck equation can be generated by a Langevin equation
\be\label{eq:langform}
m\ddot\r + m\gamma\dot\r +\partial_\r V_{\rm ext}(\r) = \bm{\eta}(\r,t)\:,
\ee
where $\bm{\eta}$ is a noise vector satisfying
\be
\avg{\bm{\eta}(\r,t)}_{\bm{\eta}} = 0\:,\qquad \avg{\eta_i(\r,t)\eta_j(\r,t')}_{\bm{\eta}} = \kappa\,\delta_{ij}\delta(t-t')\:.
\ee
The stationary solution of the Fokker-Planck equation is nothing but the Maxwell-Boltzmann equilibrium distribution \footnote{Eq.(\ref{eq:Fokker1}) admits a normalised stationary solution if and only if the integral
$
\int\!\diff{\r}\,\eu^{-\frac{V_{\rm ext}(\r)}{T}}
$
is finite. Notice that this is not the case for an attractive potential, that is $V_{\rm ext}(\r)<0$ for all $\r\in\mathbb{R}^3\,$, since it causes an instability.}
\be
\rho_{\rm eq}(\r,\p)=\mathcal{N}\,\exp\lc-\frac{1}{T}\lbr\frac{\p^2}{2m}+V_{\rm ext}(\r)\rbr\rc\:,
\ee
with the normalisation constant
\be
\mathcal{N} = \lbr\frac{\hbar^2}{2\pi mT}\,\rbr^{3/2}\lbr\int\!\diff{\r}\,\eu^{-\frac{V_{\rm ext}(\r)}{T}}\rbr^{-1}\:,
\ee
such that
\be
\int_{\r}\int_{\p}\,\rho_{\rm eq}(\r,\p)= 1\:,
\ee
which is equivalent to the requirement Tr$\hrho=1\,$.\\
The interesting case $l_{env}/l_{sys}\lesssim 1$ will be numerically analysed later for the density matrix of a $q\bar q$ pair. The last case corresponds to $l_{env}/l_{sys}\ll 1\,$. In this scenario the imaginary part of the potential has support only in a very small region\footnote{Notice that $W(|\x|\to \infty)=0$ because of the definition $W(\x)=-\int\diff{t}\,\Delta^<(t,\x)$ and the fact that the Wightman function $\Delta^<$ dies off at large space separation. Moreover $W(0)<0$ since $\Delta^<(0,0)=\avg{A_0(0,0)A_0(0,0)}>0$ and $\Delta^<(t\to\infty,0)=0\leq\Delta^<(t,0)\leq\Delta^<(0,0)$ for each time $t\,$.}. This means that, excluding the region where $\y=\q-\q'=0\,$, the Lindblad equation (\ref{eq:Lindblad1Q}) with external potential becomes 
\be\label{eq:1particlerho}
\hspace{-0.9cm}\partial_t\rho(t,\q,\q')\approx\lbr\frac{\iu\hbar}{2m}\lbr\partial_{\q}^2-\partial_{\q'}^2\rbr-\frac{\iu}{\hbar}(V_{\rm ext}(\q)-V_{\rm ext}(\q'))+\frac{g^2}{\hbar}W(0)\rbr\rho(t,\q,\q')\:.
\ee
The solution of this equation for the off-diagonal ($\q\neq\q'$) elements of the density matrix is easily found to be
\be
\rho(t,\q,\q')\approx\eu^{\,g^2\frac{W(0)}{\hbar}t}\rho_{\rm rev}(t,\q,\q')\:,
\ee
where $W(0)<0$ and $\rho_{\rm rev}(t,\q,\q')$ is the solution of the reversible von Neumann equation, i.e. $\!$eq.(\ref{eq:1particlerho}) without the dissipative term. One immediately notices that the off-diagonal elements of the density matrix become suppressed as time increases, signaling a strong quantum decoherence. In sec.\ref{sec:numerical} it will be shown numerically that the same phenomenon occurs also when $l_{\rm env}\sim l_{\rm sys}\,$, but decoherence manifests itself more slowly. This can be understood in terms of the environment performing frequent measurements on the subsystem, every time the former kicks the latter.     
When $\rho_{\rm rev}(t,\q,\q')=\avg{\q|\psi(t)}\avg{\psi(t)|\q'}$ is expanded in the basis of the Hilbert space defined by the Hamiltonian of the closed subsystem ($H\ket{\psi_n}=E_n\ket{\psi_n}$), the solution of eq.(\ref{eq:1particlerho}) reads
\be\label{eq:decay}
\rho(t,\q,\q')\approx \sum_{n,m}\eu^{\,g^2\frac{W(0)}{\hbar}t}\lambda_{n}\lambda_{m}^*\,\eu^{\frac{\iu}{\hbar}(E_m-E_n)t}\psi_n(\q)\psi^*_m(\q')\:,\quad \q\neq\q'\:,
\ee
where the quantity $\eu^{\frac{W(0)}{\hbar}t}|\lambda_{n}|^2$ is the probability associated to the $n^{\rm th}$ eigenstate of $H\,$. This means that all the bound states decay with the same rate $-g^2W(0)/\hbar=g^2T/(4\pi\hbar)\,$, which is also the rate of collisions between one heavy quark (with negligible kinetic energy) and the light quarks of the medium (see Appendix $\!$C of \cite{Beraudo:2007ky}). Therefore the environment does not discriminate between different bound states (see also \cite{Akamatsu:2011se} for a similar result). This is not surprising since, in the limit $l_{env}/l_{sys}\ll 1\,$, the large momentum kicks distributed by the medium are much larger than the typical binding energies of the bound states.\\
Following this argument, one could think that the trace of the density operator is not preserved since probabilities of the eigenstates tend to zero at large time. This reasoning does not hold since the trace is computed on the diagonal elements of the density matrix, which are not described by the solution (\ref{eq:decay}).\\
It is important to bear in mind that $l_{env}$ cannot actually be too small compared to $l_{sys}\,$. In fact, if $l_{env}$ were too small, the momentum kicks of the bath would excite modes of system that are beyond the description of this non relativistic model for the heavy probes.


\subsection{Ehrenfest relations and Langevin dynamics}

Once the semiclassical approximation has been performed, the stochastic (Langevin) dynamics can be also obtained via the Ehrenfest relations. 
It is straightforward to show that eq.(\ref{eq:Caldeira1}) is nothing but the well-known Caldeira-Leggett master equation
in the position representation \cite{1983AnPhy.149..374C,breuer2007theory}. The master equation for the density operator reads
\be\label{eq:Caldeira2}
\iu\hbar\,\deriv{\hrho(t)}{t}=\lc\hH,\hrho(t)\rc-\frac{\iu \kappa}{2\hbar}\,\lc\hat q,\lc\hat q,\hrho(t)\rc\rc+\frac{\gamma}{2}\,\lc\hat q,\left\{\hat p,\hrho(t)\right\}\rc\:,
\ee
with the friction ($\gamma$) and momentum diffusion ($\kappa$) coefficients defined in (\ref{eq:fricdiff}). From this equation  one obtains the Ehrenfest equations of motion
for the first and second moments of
the position and the momentum operators (see Appendix \ref{A2} for the derivation of the first two lines), where the hat on the operators has been omitted to simplify the
notation:
\be\label{Ehrenfest}
&&\deriv{}{t}\avg{q}=\frac{\avg{p}}{m}\:,\nn\\
&&\deriv{}{t}\avg{p}=-\avg{V_{\rm ext}'(q)}-\gamma\avg{p}\:,\nn\\
&&\deriv{}{t}\avg{q^2}=\frac{1}{m}\avg{pq+qp}\:,\nn\\
&&\deriv{}{t}\avg{pq+qp}=\frac{2}{m}\avg{p^2}-2\avg{qV_{\rm ext}'(q)}-\gamma\avg{pq+qp}\:,\nn\\
&&\deriv{}{t}\avg{p^2}=-\avg{pV_{\rm ext}'(q)+V_{\rm ext}'(q)p}-2\gamma\avg{p^2}+\kappa\:.
\ee
Observe that, due to the non-commuting nature of the operators, this set of equations is not closed for a generic potential, hence the system can not be solved in general. The Ehrenfest relations are closely related to the stochastic differential equations
\be\label{Weiner}
\diff{x}(t)&=&\frac{p(t)}{m}\diff{t}\:,\nn\\
\diff{p}(t)&=&-V_{\rm ext}'(x(t))\diff{t}-\gamma\,p(t)\diff{t}+\sqrt{\kappa}\diff{W(t)}\:,
\ee
with a white noise $\eta(t)\,$, formally given by the time derivative of the Weiner
process $W(t)\,$, i.e. $\!\deriv{W(t)}{t}=\eta(t)\:.$
The white noise satisfies $\avg{\eta(t)}_\eta=0$ and $\avg{\eta(t)\eta(t')}_\eta=\delta(t-t')\,$,
implying that the expressions in (\ref{Weiner}) are equivalent to the Langevin form (\ref{eq:langform}). It is straightforward to show that the
Ehrenfest relations (\ref{Ehrenfest}) have the same form of the noise
average of the equations of motion for the first and second moments of
the position and the momentum. In particular, the equations for the first moments,
\be
&&\left\langle\deriv{x}{t}\right\rangle_\eta=\frac{\avg{p}_\eta}{m}\:,\nn\\
&&\left\langle\deriv{p}{t}\right\rangle_\eta=-\avg{V_{\rm ext}'(x)}_\eta-\gamma\avg{p}_\eta\:,
\ee
are nothing but the Langevin equation for the Brownian motion of a particle immersed in a thermal bath.


\newpage


\section{Lindblad equation for a heavy $q\bar q$ pair}\label{sec:2Q}

The same procedure used for the single particle case is applied here to the case of a heavy $q\bar q$ pair propagating through the plasma. The interaction part of the influence functional appearing in the path integral (\ref{pathint}) is derived from eq.(\ref{eq:QbarQ}) for $N=2$ and reads
\be
\!\!\!\!\!\!\!\!\!\!\!\!\!\!\!\!\!\!\!\!\!&&\Phi_{\rm int}[\q,\q';\bar\q,\bar\q'] =-\int_{t_0}^{t} \diff{\tau}
\left[\vphantom{\pderiv{W(\bar\q-\q')}{\bar\q}} V(\q-\bar\q)-V(\q'-\bar\q')\right.\nn\\
&&-\iu W(\q'-\bar\q')-\iu W(\q-\bar\q)+\iu W(\q-\overline\q')+\iu W(\q'-\bar\q)\nn\\
&&+\left. \frac{\beta\hbar}{4}\left(\pderiv{W(\q-\bar\q')}{\bar\q'}\cdot(\dot{\bar\q}'+\dot\q)- \pderiv{W(\bar\q-\q')}{\bar\q}\cdot(\dot{\bar\q}+\dot\q')\right)\right]\:,
\ee
where the coordinates $(\q,\q')$ and $(\bar \q,\bar \q')$ refer to the heavy quark and antiquark respectively.
The non-interacting term for the single quark is given again by eq.(\ref{eq:singleQ}) and analogously for the antiquark with the replacements $\q\to\bar\q$ and $\q'\to\bar\q'$. As before, the phase-space path integral for the conditional probability will be derived. For this purpose, one again makes use of the identity (\ref{eq:gaussidentity}), this time written in the form
\be\label{eq:gaussidentity2}
&&\int\diff{p}\int\diff{p'} \exp\lc \frac{\iu}{\hbar}\lbr \frac{p'^2}{2m}-\frac{p^2}{2m}+ p f(q,q',\bar q',\dot q)- p'h(q,q',\bar q,\dot q') \rbr \rc \nn\\
&=& (2\pi m\hbar)\exp\lc \frac{\iu m}{2\hbar}\lbr f^2(q,q',\bar q',\dot q)-h^2(q,q',\bar q,\dot q') \rbr\rc\:,
\ee
with
\be
&&f(q,q',\bar q',\dot q)=\dot q +\frac{\beta\hbar}{4m}\lbr\partial_{q'} W(q-q')-\partial_{\bar q'}W(q-\bar q')\rbr\:,\nn\\
&&h(q,q',\bar q,\dot q')=\dot q' - \frac{\beta\hbar}{4m}\lbr\partial_{q'} W(q-q')+\partial_{\bar q}W(\bar q-q')\rbr\:,
\ee
and similarly for the heavy antiquark with the obvious replacements $\q\to\bar\q$ and $\q'\to\bar\q'$.
Neglecting terms of order $g^4$ or higher, the propagator in phase space becomes
\be\label{eq:P2Q}
&&\hspace{-1cm}P(\q_f,\q'_f,\bar\q_f,\bar\q'_f, t|\q_0,\q_0',\bar\q_0,\bar\q_0', t_0)=\!\!\int_{(\q_0,t_0)}^{(\q_f,t)}\!\!\!\!\!\!\!\!\!\!\D \q\D \p\!\!\int_{(\q_0',t_0)}^{(\q'_f,t)}\!\!\!\!\!\!\!\!\!\!\D \q'\D \p'\!\!\int_{(\bar\q_0,t_0)}^{(\bar\q_f,t)}\!\!\!\!\!\!\!\!\!\!\D \bar\q\D \bar\p\!\!\int_{(\bar\q_0',t_0)}^{(\bar\q'_f,t)}\!\!\!\!\!\!\!\!\!\!\D \bar\q'\D \bar\p'\nn\\
&&\hspace{-1cm}\exp\!\!\lc \frac{\iu}{\hbar}\!\lbr S_{\rm cl}[\q,\p,\bar\q,\bar\p] -S_{\rm cl}[\q',\p',\bar\q',\bar\p']\rbr\! \rc\!\! \F[\q,\p,\bar\q,\bar\p;\q',\p',\bar\q',\bar\p']\:,
\ee
with the classical action
\be
S_{\rm cl}[\q,\p,\bar\q,\bar\p]= \int_{t_0}^t\diff{\tau}\lbr \dot \q\p+\dot{\bar\q}\bar\p-\frac{\p^2}{2m}-\frac{{\bar\p}^2}{2m}-V(\q-\bar\q)\rbr\:,
\ee
and the in-medium contribution\footnote{Notice that the line containing the second derivatives of $W$ is a negligible o$(g^4)$ term that has been added by hand in order to exactly preserve the trace of the density operator. The same procedure was carried out for a single quark (see eq.(\ref{eq:negl})).}
\be
&&\F[\q,\p,\bar\q,\bar\p;\q',\p',\bar\q',\bar\p']\nn\\
&&=\exp\lc \frac{\iu}{\hbar}g^2\!\!\int_{t_0}^{t}\!\!\diff{\tau}\!\lc\frac{}{} \iu(W(\q-\q')+W(\bar\q-\bar\q')+W(\q-\bar\q)\right.\right.\nn\\
&&\left.\left.+W(\q'-\bar\q') - W(\q-\overline\q')- W(\q'-\bar\q)-2W(0))\right.\right.\nn\\
&&\left.\left. -\frac{\iu\beta\hbar^2}{2m}\lbr \pderiv[2]{W(\q-\q')}{\q} +\pderiv[2]{W(\bar\q-\bar\q')}{\bar\q} \rbr  \right.\right.\nn\\
&&\left.\left. + \frac{\beta\hbar}{4m}\lbr\pderiv{W(\q-\q')}{\q'}\cdot(\p+\p')+\pderiv{W(\bar\q-\bar\q')}{\bar\q'}\cdot(\bar\p+\bar\p')\right.\right.\right.\nn\\
&&\left.\left.\left.+\pderiv{W(\q-\bar\q')}{\q}\cdot(\p+\bar\p') +\pderiv{W(\bar\q-\q')}{\bar\q}\cdot(\bar\p+\p')\rbr \rc\!\!+\mbox{o}(g^4)\rc\:.
\ee
From the last expression and definition (\ref{eq:F}), one reads off the phase functional and distance decoherence functional for the $q\bar q$ pair:
\be\label{eq:varphi2Q}
\varphi &=& \frac{g^2\beta\hbar}{4m}\int_{t_0}^{t}\!\!\diff{\tau}\lbr\pderiv{W(\q-\q')}{\q'}\cdot(\p+\p')+\pderiv{W(\bar\q-\bar\q')}{\bar\q'}\cdot(\bar\p+\bar\p')\right.\nn\\
&&\left. -\pderiv{W(\q-\bar\q')}{\bar\q'}\cdot(\p+\bar\p') +\pderiv{W(\bar\q-\q')}{\bar\q}\cdot(\bar\p+\p') \rbr\:,
\ee
and
\be\label{eq:D2Q}
&&\!\!\!\!\!\!\!\!D^2= 2g^2\int_{t_0}^{t}\!\!\diff{\tau}\lc\frac{}{}(W(\q-\q')+W(\bar\q-\bar\q') +W(\q'-\bar\q')+W(\q-\bar\q)\right.\\
&&\!\!\!\!\!\!\!\!\left.-W(\q-\bar\q')-W(\q'-\bar\q)-2W(0))-\frac{\beta\hbar^2}{4m}\lbr\pderiv[2]{W(\q-\q')}{\q} +\pderiv[2]{W(\bar\q-\bar\q')}{\bar\q} \rbr\rc\:.\nn
\ee
Observe that $\varphi$ vanishes in the coincidence limit $(\q',\p',\bar\q',\bar\p')=(\q,\p,\bar\q,\bar\p)$, and $D^2$ is symmetric when $(\q',\p',\bar\q',\bar\p')$ and $(\q,\p,\bar\q,\bar\p)$ are swapped, as required by the definitions (\ref{eq:varphi}) and (\ref{eq:D}).
Following a very similar calculation to the one performed in sec.\ref{sec:1Q} for one heavy quark, one finds the Lindblad equation for a $q\bar q$ pair in the plasma\footnote{Notice that the Lindblad equation is symmetric under quark-antiquark exchange, as physically expected}:
\be\label{eq:Lindblad2Q}
&&\!\!\!\!\!\!\partial_t\rho(t,\q,\q',\bar\q,\bar\q')=\lbr\frac{\iu\hbar}{2m}\lbr\partial^2_{\q}-\partial^2_{\q'}+\partial^2_{\bar\q}-\partial^2_{\bar\q'}\rbr-\frac{\iu g^2}{\hbar}\lbr\frac{}{}V(\q-\bar\q)-V(\q'-\bar\q')\frac{}{}\rbr\right.\nn\\
&&\!\!\!\!\!\!\left.-\frac{g^2}{\hbar}\lbr\frac{}{}W(\q-\q')+W(\bar\q-\bar\q')+W(\q'-\bar\q')+W(\q-\bar\q)\right.\right.\nn\\
&&\!\!\!\!\!\!\left.\left.-W(\q-\bar\q')-W(\bar\q-\q')-2W(0)\frac{}{}\rbr-\frac{g^2\beta\hbar}{4m}\lc\frac{}{}\partial_{\q'} W(\q-\q')\cdot\lbr \partial_{\q'}-\partial_{\q} \rbr\right.\right.\nn\\
&&\!\!\!\!\!\!\left.\left.+\partial_{\bar\q'} W(\bar\q-\bar\q')\cdot\lbr \partial_{\bar\q'}-\partial_{\bar\q} \rbr-\partial_{\bar\q'} W(\q-\bar\q')\cdot(\partial_{\bar\q'}-\partial_\q)+\partial_{\bar\q} W(\bar\q-\q')\cdot(\partial_{\q'}-\partial_{\bar\q})\right.\right.\nn\\
&&\!\!\!\!\!\!\left.\left.-\partial_{\bar\q'}^2W(\q-\bar\q')-\partial_{\q'}^2W(\bar\q-\q')\frac{}{}\rc \rbr \rho(t,\q,\q',\bar\q,\bar\q')\:.
\ee
An interesting feature can be inferred from this result.
If the quark and antiquark are taken far apart, which implies that $|\q-\bar\q|,|\q'-\bar\q'|\gg r_{_D}\,$, where $ r_{_D}$ is the Debye radius, and consider that\footnote{The distances $|\bar\q-\bar\q'|$ and $|\bar\q'-\bar\q|$ are of the order of the de Broglie wavelength of the heavy particles.}
\be
&&|\q-\bar\q'|\geq |\q-\bar\q|-|\bar\q-\bar\q'|\approx |\q-\bar\q| \gg r_{_D}\:,\nn\\
&&|\q'-\bar\q|\geq |\q'-\bar\q'|-|\bar\q'-\bar\q|\approx |\q'-\bar\q'| \gg r_{_D}\:,
\ee
one finds that eq.(\ref{eq:Lindblad2Q}) admits a factorised solution
\be
\rho(t,\q,\q',\bar\q,\bar\q')=\rho_q(t,\q,\q')\rho_{\bar q}(t,\bar\q,\bar\q')\:,
\ee
since both the real and imaginary part of the potential vanish when their arguments are large compared to the Debye radius. The density matrix for the single heavy particle satisfies the Lindblad equation (\ref{eq:Lindblad1Q}).
Eventually, after an integration by parts, it is immediate to check that the Lindbald equation (\ref{eq:Lindblad2Q}) preserves the trace, that is
\be
\deriv{}{t}\int_{\q}\int_{\bar\q}\int_{\q'}\int_{\bar\q'}\,\delta(\q-\q')\delta(\bar\q-\bar\q')\rho(t,\q,\q',\bar\q,\bar\q')=0\:.
\ee


\subsection{Tracing out the motion of the centre of mass}

Since the motion of the centre of mass of the $q\bar q$ pair does not tell anything about the fate of bound states, one would like to trace out this collective motion and find a Lindblad equation for the density matrix describing only the dynamics of the relative $q\bar q$ coordinate.
To accomplish this, one starts off by defining the centre of mass and relative coordinates respectively as
\be
&&\q_{\rm cm}\equiv\frac{1}{2}(\q+\bar\q)\:,\nn\\
&&\q_r \equiv \q-\bar\q\:,
\ee 
and similarly for the primed coordinates. The density matrix describing the dynamics of the relative $q\bar q$ coordinate is defined as the trace of the density matrix with respect to the centre of mass coordinate, that is
\be
\tilde\rho(t,\q_r,\q_r')=\int_{\q_{\rm cm}}\int_{\q_{\rm cm}'}\!\!\delta(\q_{\rm cm}-\q_{\rm cm}')\rho(t,\q_r,\q_{\rm cm},\q_r',\q_{\rm cm}')\:.
\ee
In the coincident limit $\q_{\rm cm}'\to\q_{\rm cm}\,$, the Lindblad operator does not depend on $\q_{\rm cm}$ anymore, therefore one can carry out the trace over the centre of mass coordinate and get the Lindblad equation for the relative motion:
\be
&&\!\!\!\!\!\!\!\!\partial_t\tilde\rho(t,\q_r,\q_r')=\lc \frac{\iu\hbar}{m}(\partial_{\q_r}^2-\partial_{\q_r'}^2) -\frac{\iu g^2}{\hbar}\lbr V(\q_r)-V(\q_r') \rbr   \right.\\
&&\!\!\!\!\!\!\!\!\left. -\frac{g^2}{\hbar}\lbr 2W\lbr\frac{1}{2}(\q_r-\q_r')\rbr -2W\lbr\frac{1}{2}(\q_r+\q_r')\rbr +W(\q_r)+W(\q_r')-2W(0)\rbr \right.\nn\\
&&\!\!\!\!\!\!\!\!\left.  -\frac{g^2\beta\hbar}{m}\lbr \partial_{\q_r}W\lbr\frac{1}{2}(\q_r-\q_r')\rbr\cdot(\partial_{\q_r}-\partial_{\q_r'})-\partial_{\q_r}W\lbr\frac{1}{2}(\q_r+\q_r')\rbr\cdot(\partial_{\q_r}+\partial_{\q_r'})\right.\right.\nn\\
&&\left.\left. -2\partial_{\q_r}^2W\lbr\frac{1}{2}(\q_r+\q_r')\rbr  \rbr \rc\tilde\rho(t,\q_r,\q_r')\:.\nn
\ee
Performing a change of variables analogous to (\ref{eq:change}) but with $\y$ halved, that is
\be\label{eq:change2}
\r\equiv\frac{1}{2}(\q_r+\q_r')\:,\qquad \y \equiv\frac{1}{2}(\q_r-\q_r')\:,
\ee
one finally obtains
\be\label{eq:Linddim}
&&\!\!\!\!\!\!\!\!\partial_t\tilde\rho(t,\r,\y)=\lc \frac{\iu\hbar}{m}\partial_\r\cdot\partial_\y -\frac{\iu g^2}{\hbar}\lbr V(\r+\y)-V(\r-\y) \rbr\right.\\
&&\!\!\!\!\!\!\!\!\left. -\frac{g^2}{\hbar}\lbr 2W(\y) -2W\lbr\r\rbr +W(\r+\y)+W(\r-\y)-2W(0)\rbr \right.\nn\\
&&\!\!\!\!\!\!\!\!\left.  -\frac{g^2\beta\hbar}{2m}\lbr \partial_{\y}W(\y)\cdot\partial_{\y}-\partial_{\r}W\lbr\r\rbr\cdot\partial_{\r}-\partial_{\r}^2W(\r)  \rbr \rc\tilde\rho(t,\r,\y)\:.\nn
\ee
Observe that, as in the one particle case, the diffusive term (second line of (\ref{eq:Linddim})), responsible for quantum decoherence, affects only the off-diagonal elements of the density matrix. 
Moreover, using integration by parts, it is immediate to check that the trace of the density matrix is preserved.


\subsection{Fokker-Planck and Langevin dynamics in the classical limit}

In section \ref{sec:sc1} it has been shown that the semiclassical limit of the master equation for a single heavy particle produces a classical stochastic dynamics. The same happens when one takes the semiclassical limit of eq.(\ref{eq:Linddim}) by expanding up to $\y^2$ (the expansion is actually in the dimensionless quantity $\y/l_{\rm env}$). The result is
\be\label{eq:Lind2sc}
&&\!\!\!\!\!\!\!\!\partial_t\tilde\rho(t,\r,\y)=\lc \frac{\iu\hbar}{m}\partial_\r\cdot\partial_\y -\frac{2\iu g^2}{\hbar}\y\cdot\partial_{\r} V(\r) -\frac{g^2}{\hbar}\y\cdot\lbr \h(0)+\h(\r)\rbr\cdot\y\right.\\
&&\!\!\!\!\!\!\!\!\left.  -\frac{g^2\beta\hbar}{2m}\lbr \y\cdot\h(0)\cdot\partial_{\y}-\partial_{\r}W\lbr\r\rbr\cdot\partial_{\r}-\partial_{\r}^2W(\r)  \rbr \rc\tilde\rho(t,\r,\y)\:,\nn
\ee
with the Hessian matrix $\h(\r)$ defined through $\h(\r)_{ij}=\partial^2W(\r)/\partial r_i\partial r_j\,$. Taking the Wigner transform\footnote{Notice the extra factors of $2$ in the definition of the Wigner function. They stem from the fact that a halved $\y$ coordinate has been used in the change of variables (\ref{eq:change2}).}
\be
\tilde\rho(t,\r,\p)=2^d\!\int_{\y}\,\tilde\rho(t,\r,\y)\,\eu^{-\frac{2\iu}{\hbar}\p\cdot\y}
\ee
of eq.(\ref{eq:Lind2sc}), one finds the following Fokker-Planck-like equation for the Wigner function:
\be\label{eq:Fokkermulti}
&&\!\!\!\lc \partial_{t} + \lbr \frac{2\p}{m}-\frac{g^2\hbar\beta}{2m}\partial_\r W(\r) \rbr\!\cdot\partial_\r -g^2\partial_\r V(\r)\cdot\partial_\p\rc\tilde\rho(t,\r,\p)=\\
&&\!\!\!\frac{g^2\hbar}{4}\lc \partial_\p\!\cdot\!\lbr \h(0)+\h(\r) \rbr\!\cdot\!\partial_\p +\frac{2\beta}{m}\lbr \left.\partial_\r^2W(\r)\right|_{\r=\bold{0}} + \partial_\r^2W(\r) +\p\!\cdot\!\h(0)\!\cdot\partial_\p \rbr \rc\tilde\rho(t,\r,\p)\:.\nn
\ee
As observed in section \ref{sec:sc1}, this becomes a proper Fokker-Planck equation when the Wigner function turns into a classical probability distribution in phase space. This transition happens on time scale of the order of the decoherence time, which is much smaller than the typical relaxation time of the subsystem.
Appendix \ref{A1} shows how equation (\ref{eq:Fokkermulti}) can be obtained from the following Langevin equation for the relative motion of the $q\bar q$ pair, with position-dependent friction and noise\footnote{Also known as \textit{multiplicative} noise.} (see also Sec. $\!$4.3 of \cite{Blaizot:2015hya}):
\be\label{eq:Langmulti}
\mu\ddot\r + \mu\gamma(\r)\cdot\dot\r + g^2\partial_\r V(\r)= \boldeta(\r,t)\:,
\ee
where $\mu=m/2$ is the reduced mass of the subsystem and the friction matrix is
\be
\gamma(\r) \equiv \frac{ g^2\hbar}{4\mu T}\lbr \h(0)+\h(\r) \rbr\:,\nn\\
\ee
The multiplicative noise $\boldeta(\r,t)$ is white and autocorrelated, being
\be
\avg{\boldeta(\r,t)}_{\boldeta}=\bm{0}\:,\qquad \avg{\eta_i(\r,t)\eta_j(\r,t')}_{\boldeta}=\kappa_{ij}(\r)\delta(t-t')\:,
\ee
where the momentum diffusion matrix
\be\label{eq:difflang}
\kappa(\r) \equiv \frac{g^2\hbar}{2}\lbr \h(0)+\h(\r) \rbr\:,
\ee
satisfies the Einstein relation $\kappa(\r)=2\mu\gamma(\r)T$ required by the \textit{fluctuation-dissipation} theorem applied to the Brownian motion of a particle in a thermal bath \cite{schwabl2006statistical}.\\
Notice that a Langevin equation is much easier to be evaluated numerically than a Fokker-Planck equation. However, the information about the quantum $q\bar q$ bound states at $t=0$ can not be encoded in the initial conditions of a Langevin equation, which is classical. On the other hand, the quantum initial conditions are easily implemented in the Fokker-Planck-like equation for the Wigner function. One could think of the Langevin equation as the correct formalism to describe the dynamics of the $q\bar q$ bound states after a typical time scale of the order of the decoherence time, which is much shorter than the typical relaxation time of the subsystem when $l_{\rm env}\lesssim l_{\rm sys}$, as it will be displayed in sec.\ref{sec:numerical}.


\section{1D simulation for a heavy $q\bar q$ pair}\label{sec:modelsim}

The target of this section is to analyse dissociation, (re)combination  and quantum decoherence
of bound states of a heavy $q\bar q$ pair in a medium. To achieve this, the Lindblad equation (\ref{eq:Linddim}) will be solved numerically using Mathematica.\\
This part presents a simple model to investigate numerically the effect of the Lindbladian terms on the dynamics and fate of quarkonia. These terms depend on the imaginary part of the potential ($W$), which originates from the Landau mechanism due to the collisions between the subsystem and the medium. The real part of the potential ($V$) is found instead in the reversible part of the Lindblad equation. Surely the screening of the real part of the potential reduces the number of available bound states in the system, however, this is not the effect that this work aims to focus on. The screening mechanism is important when one wants to consider phenomenological implications, but this is not the case of this exploratory analysis. Here the focus is on understanding how the imaginary part of the potential influences the dynamics of the quantum system.


\subsection{Bound-state probabilities and linear entropy}

The solution of the Lindblad equation is used to compute the following two quantities for studying bound-state dissociation/formation and decoherence respectively:
\begin{itemize}
\item $P(\psi, t|\psi_0, t_0)=\int_{q}\!\int_{q'}\psi(q')\psi^*(q)\rho(t,q,q')\:$;
\item $S_{_L}=\mbox{Tr}\hrho-\mbox{Tr}\hrho^2=1-\mbox{Tr}\hrho^2\:$.
\end{itemize}
The first quantity was introduced previously in eq.(\ref{eq:P1}) and represents the probability of finding the subsystem in the state $\ket{\psi}$ at time $t\,$, given that it was origianally in the state $\ket{\psi_0}$ at $t=0\,$\footnote{The initial condition has to be specified when solving for the differential equation of the density matrix.}. The second object is the so-called linear entropy, which is the leading term of the Mercator series of the von Neumann entropy $S=-\mbox{Tr}[\hrho\ln\hrho]\approx \mbox{Tr}[\hrho(\Id-\hrho)]=S_{_L}\,$. The computation of the standard entropy requires the diagonalisation of the density matrix, which can be done only when all the probabilities $\{P_n(t)\}_n$ of the eigenstates $\{\psi_n\}_n$ of the Hilbert space of the subsystem are known. This is usually an impossible task, this is why the linear entropy is computed as a proxy of the von Neumann one.\\
Another reason to consider the linear entropy is that it is a handy quantity for diagnosing quantum decoherence. Indeed a pure state has $S_{_L}=0\,$, being $\ket{\psi(t)}\bra{\psi(t)}=\hrho(t)=\hrho^2(t)\,$, whereas a mixed state has $0<S_{_L}\leq 1\,$, since
\be
0\leq\mbox{Tr}\,\hrho_{\rm mix}^2=\sum_n P_n^2 \leq \sum_n P_n=1\:;\qquad \hrho_{\rm mix}=\sum_nP_n\ket{\psi_n}\bra{\psi_n}\:.
\ee
If a pure state of the subsystem is plunged into the bath, the linear entropy will increase with a rapidity dependent on the ability of the medium to resolve the quantum system, that is according to the ratio $l_{\rm env}/l_{\rm sys}$ that has been discussed in section \ref{sec:sc1}. If this ratio is very small, the environment performs ``invasive'' measures on the subsystem (the momentum kicks of the bath are large), inducing a sudden raise of the linear entropy. In the opposite regime, the environment can not resolve the system and consequently the linear entropy remains close to zero.\\
It is interesting to see how the off-diagonal elements of the density matrix affect the time evolution of the linear entropy in the semiclassical limit ($l_{\rm env}\gtrsim l_{\rm sys}$). Using that $\dot S_{_L} = -2\,\mbox{Tr}\lc\dot\hrho\hrho\rc$ and inserting eq.(\ref{eq:Lind2sc}), after some algebra one gets
\be
\dot S_{_L}(t) \approx \frac{8\kappa}{\hbar^2}\int_{\r}\int_{\y}\,\y^2|\rho(t,\r,\y)|^2\geq 0\:,
\ee
where it has been used that $\left.\partial_\r^2W(\r)\right|_{\r=\bold{0}} - \partial_\r^2W(\r)\approx 0$ when $l_{\rm env}\gtrsim l_{\rm sys}\,$. It is clear that in this regime the linear entropy rises as long as the density matrix has off-diagonal elements. This makes the role of off-diagonal terms in the process of quantum decoherence more transparent.


\subsection{Model for simulations}

\begin{figure}[t!]
\begin{center}
\includegraphics[width=8.5cm]{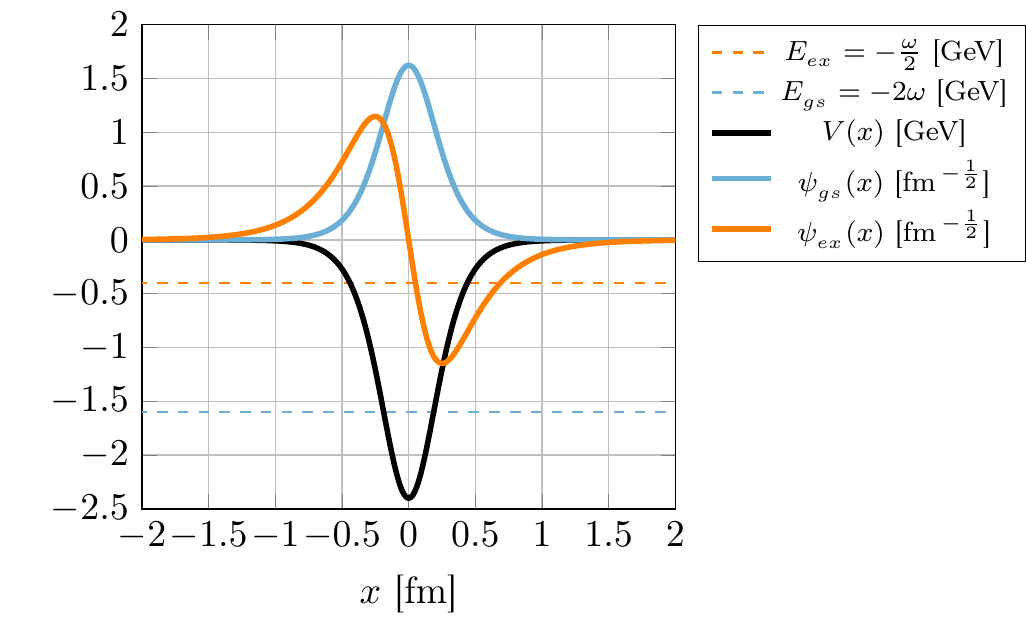}
\includegraphics[width=6cm]{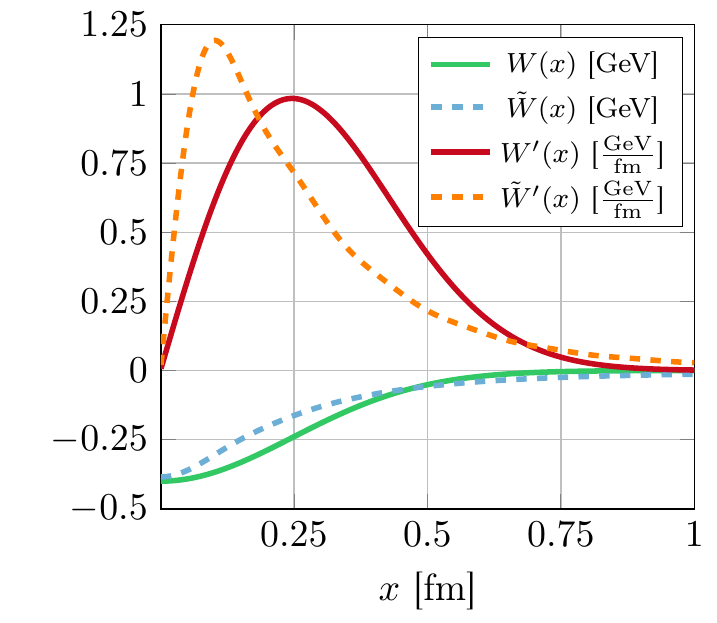}
\caption{On the left: P\"oschl-Teller potential with $\omega=0.8$ GeV, $\mu=0.6$ GeV and $j=2$ bound states. The ground state $\psi_{_{\rm gs}}$ and first excited state $\psi_{_{\rm ex}}$ with their energies are also shown. On the right: Imaginary part of the potential (\ref{eq:im}) used in the simulations contrasted with $\tilde W(x)$ derived from the gauge theory in (\ref{W}). Here $T=0.8$ GeV and $\sigma=1\,$, hence $l_{\rm env}=\hbar/(\sigma T)\approx 0.25$ fm.}
\label{fig:Potentials}
\end{center}
\end{figure}

Instead of using the expressions in (\ref{W}) for $V$ and $W\,$, two similar expressions are used here for computational convenience. The real part is taken to be the P\"oschl-Teller potential (see Fig.\ref{fig:Potentials})
\be\label{eq:Poschl}
V(x)=-\frac{\omega}{2}j(j+1)\sech^2\lc \sqrt{\frac{\mu\,\omega}{\hbar^2}}x \rc\:,
\ee 
where $\omega$ is a unit of energy, $\mu=m/2$ is the reduced mass of the $q\bar q$ pair and $j\in\mathbb{N}\,$. This potential has been chosen since it admits $j$ bound states,
\be\label{eq:2wavefunctions}
\psi_n(x)=C_n^j\lbr\frac{\mu\,\omega}{\hbar^2}\rbr^{1/4}\mbox{P}_j^n\lbr\tanh\lc\sqrt {\frac{\mu\,\omega}{\hbar^2}}x\rc\rbr\;,
\ee
with $n=0,1,\dots,j-1,j\,$, $C_n^j$ a dimensionless normalisation factor and $\mbox{P}_j^n(x)$ the associated Legendre polynomial.
Therefore the maximum number of vacuum ($T=0$) bound states can be simply tuned by changing the natural number $j\,$, which is set to $2$ here. The imaginary part is taken to be (see Fig.\ref{fig:Potentials})
\be\label{eq:im}
W(x)=-\frac{T}{2}\,\exp\lc-\frac{1}{2}\lbr\frac{x}{l_{\rm env}}\rbr^2\rc\:,
\ee
where the correlation length of the environment is $l_{\rm env}=\hbar/(\sigma\,T)$ and $\sigma$ is a dimensionless parameter that will be tuned to probe the different physical regimes discussed in section \ref{sec:sc1}. This form of the correlation length resembles the one of the Debye radius $r_{_D}=m_{_D}^{-1}=\hbar/(\alpha g\,T)$ appearing in the original expression (\ref{W}) of $W\,$, which is plotted in Fig.\ref{fig:Potentials} as $\tilde W(x)=-\frac{Tl_{\rm env}}{|x|}\int_0^5\diff{z}\frac{\sin\lc\frac{2z|x|}{l_{\rm env}}\rc}{(z^2+1)^2}\,$. Comparing with the expression derived from the gauge theory, a factor $\frac{1}{4\pi}$ has been omitted in front of $\tilde W(x)\,$. This has been done since the model is one dimensional anyway, and the form (\ref{W}) is strictly valid only in three dimensions.
In the numerical simulations, the bound-state probabilities ($P_0(t)$ and $P_1(t)$ for the ground and excited state respectively) and the linear entropy will be extracted from the solution of the following one-dimensional Lindblad equation for the relative motion of a $q\bar q$ pair:
\be\label{eq:LindMath}
&&\!\!\!\!\!\!\!\!\partial_t\rho(t,r,y)=\lc \frac{\iu\hbar}{m}\partial_r\partial_y -\frac{\iu}{\hbar}\lbr V(r+y)-V(r-y) \rbr\right.\\
&&\!\!\!\!\!\!\!\!\left. -\frac{1}{\hbar}\lbr 2W(y) -2W\lbr r\rbr +W(r+y)+W(r-y)-2W(0)\rbr \right.\nn\\
&&\!\!\!\!\!\!\!\!\left.  -\frac{\hbar}{2mT}\lbr \partial_{y}W(y)\partial_{y}-\partial_{r}W\lbr r\rbr\partial_{r}-\partial_{r}^2W(r)  \rbr \rc\rho(t,r,y)\:.\nn
\ee
This equation is the one-dimensional analog of eq.(\ref{eq:Linddim}) with $g^2=1$ \footnote{Notice that the perturbative approximation is still valid since the expansion is actually in powers of $\alpha_s=\frac{g^2}{4\pi}\,$. Moreover $T=0.8$ GeV and $m=1.2$ GeV have been used here, so that the nonrelativistic limit is valid to a good degree of accuracy.} and $V(x),\,W(x)$ defined as in (\ref{eq:Poschl}),(\ref{eq:im}) respectively. The first part of the analysis assumes that initially the $q\bar q$ pair is in one of the two bound states of the P\"oschl-Teller potential, that is
\be
\rho(t=0,r,y)=\psi_n(r+y)\psi_n^*(r-y)\:,
\ee
where $n=0,1$ for the ground state and excited state respectively. If the bound states melt, the density matrix describes scattering states, whose density matrix does not go to zero at infinity. To avoid this numerical issue, the system is inserted in a harmonic box via the following modification of the real part of the potential\footnote{The bound-state probabilities and linear entropy computed in the following section do not depend on the chosen value of $\epsilon$, as long as it is small enough to allow the ground and excited states of $V(x)$ to be eigenstates also of the modified potential $V(x)+\epsilon x^2\,$.}:
\be\label{eq:unbounded}
V(x)\to V(x) + \epsilon\,x^2\:,
\ee
with $\epsilon$ very small, so that the P\"oschl-Teller potential gets significantly modified only for $|x|\gtrsim 2$ fm, where the wavefunctions of the bound states are basically zero. Therefore, for numerical purposes, the eigenstates of the P\"oschl-Teller potential are also eigenstates of the P\"oschl-Teller potential plus the harmonic contribution. In addition, the new potential admits an infinite set of more energetic bound states, which can be interpreted as localized ``scattering'' states of the unmodified P\"oschl-Teller potential. This trick allows one to use vanishing boundary conditions for the density matrix when solving (\ref{eq:LindMath}) in a finite box.\\
In the cases analysed below, it has been checked numerically that the density operator is positive-semidefinite, that is $\rho(t,q,q)\geq 0$ for all times $t$ and positions $q$ (equivalently $\rho(t,r,y=0)\geq 0$ $\forall t,r\,\in\mathbb{R}\,$), as required by a master equation in the Lindblad form. Moreover, eq.(\ref{eq:LindMath}) preserves the trace of the density operator to the value of one with an error of the order of one part in a thousand.


\section{Numerical results}\label{sec:numerical}


\subsection{Dissociation of bound states}

\begin{figure}[t!]
\begin{center}
\includegraphics[width=8.5cm]{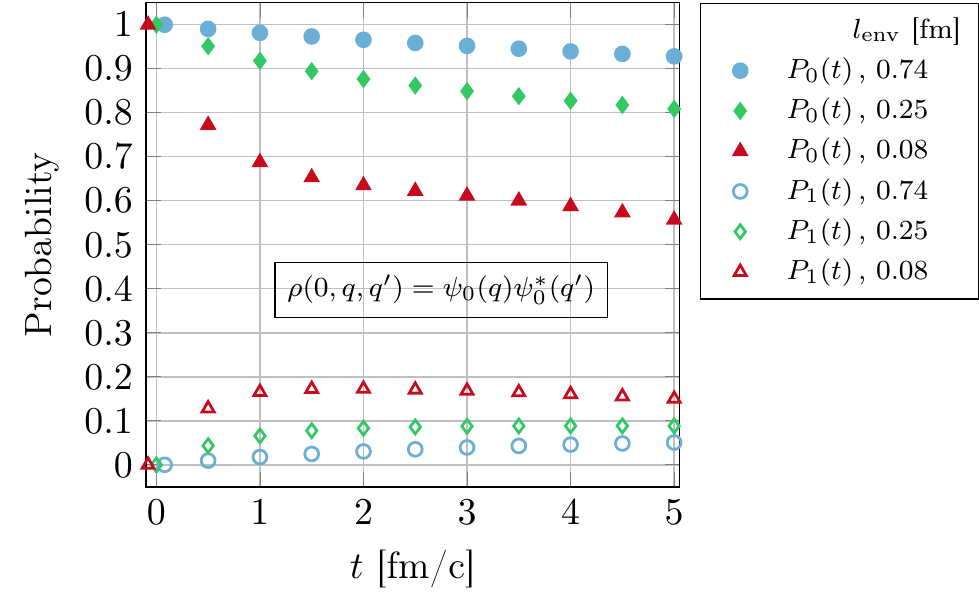}
\includegraphics[width=6cm]{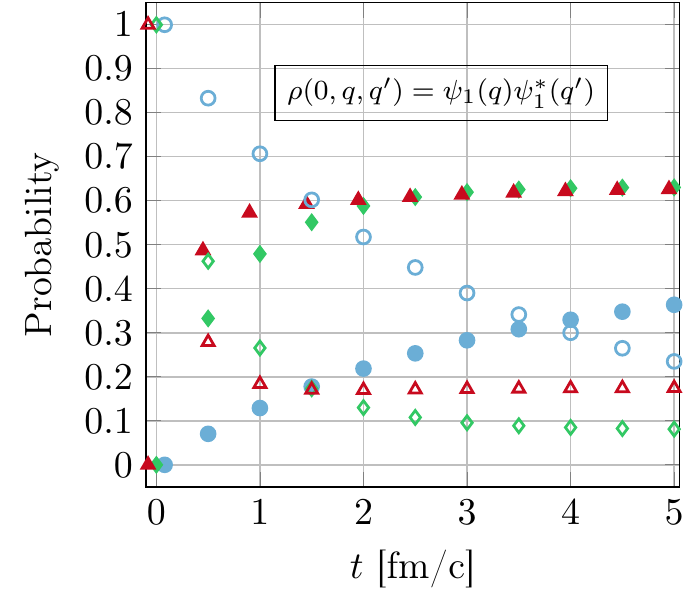}
\includegraphics[width=8cm]{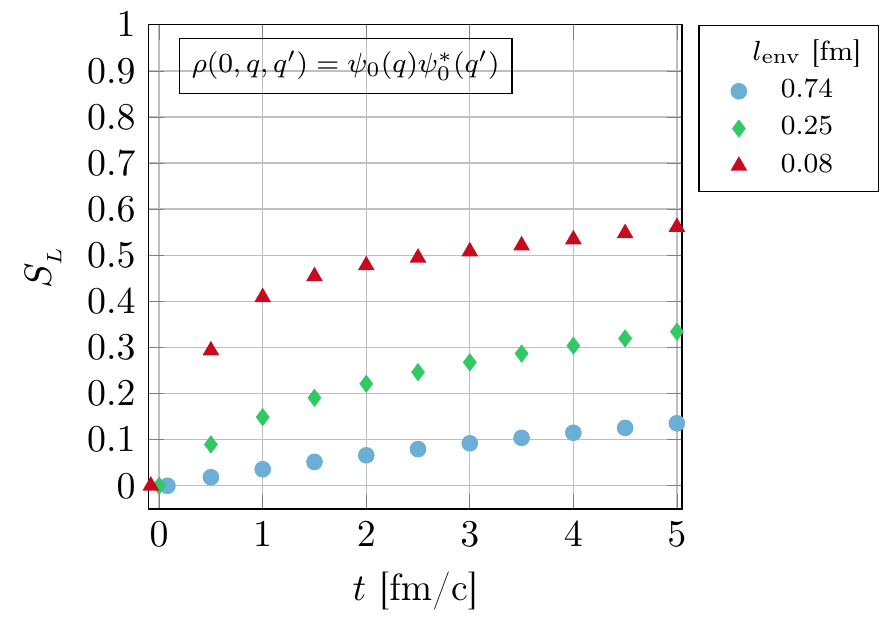}
\includegraphics[width=6.3cm]{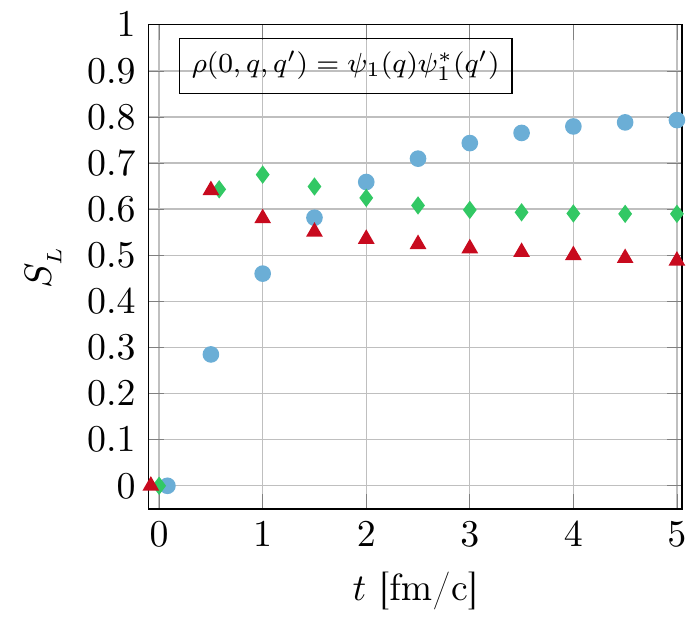}
\end{center}
\caption{Top: Probabilities $P_0(t)\,,\,P_1(t)$ of having respectively the ground state $\psi_0$ and the excited state $\psi_1$ at time $t\,$. Bottom: Time evolution of the linear entropy. In the left (right) panels the initial density matrix $\rho(0,q,q')$ corresponds to the ground (excited) state. Three different correlation lengths of the environment ($l_{\rm env}$) have been considered in order to explore the regimes $l_{\rm env}>l_{\psi_1}\,$, $l_{\psi_0}<l_{\rm env}<l_{\psi_1}\,$, $l_{\rm env}<l_{\psi_0}\,$, where $l_{\psi}=\sqrt{\avg{x^2}}_{\psi}$ with values $l_{\psi_0}=0.162$ fm and $l_{\psi_1}=0.384$ fm. In all plots of this paper error bars are much smaller than the symbols representing the data. Some of the points are slightly shifted horizontally to avoid superpositions of symbols.}
\label{fig:probab-entropy}
\end{figure}

The probabilities of having one of the two bound states at time $t\,$, given an initial density matrix corresponding to the ground (excited) state, are shown in the top left (right) panel of Fig.\ref{fig:probab-entropy}. The time evolution of the linear entropy for the two different initial density matrices is shown just in the corresponding panels below.
The analysis of the outcomes is divided in three cases, according to the correlation length of the environment.
\begin{itemize}
\item $l_{\rm env}> l_{\psi_1}$ regime
\end{itemize}
In this scenario the correlation length of the bath is bigger than the size of the excited state, which in turns is larger than the size of the ground state.
This means that the environment can ``measure'' the excited state but can very hardly resolve the ground state. This has indeed been found on the left panels of Fig.\ref{fig:probab-entropy}, where the initial ground state (see circles) remains basically unaffected as time passes ($P_0(t)\gtrsim 90\%$ $\forall t\,$). On the other hand, there is a small probability of having a transition from the ground state to the excited state($P_1(t)\lesssim 10\%$ $\forall t\,$), but there is basically no change of dissociation since $P_0(t)+P_1(t)\approx 1=P_0(t)+P_1(t)+P_{\rm scatt}(t)$ at all times, where $P_{\rm scatt}$ is the probability of having some ``scattering'' states.
Not only the medium cannot give large enough momentum kicks to disturb the ground state, but also cannot resolve it, hence quantum decoherence should manifest itself as a very mild effect. This is indeed seen in the small values taken by the linear entropy, indicating that the initial state remains approximately pure.\\
When starting off with the excited state (circles on the right panels of Fig.\ref{fig:probab-entropy}), the linear entropy increases quite rapidly up to a stationary value of $S_{_L}\approx 0.8\,$, meaning that the measurements that medium makes on the subsystem bring relatively quickly the excited state from a pure to a mixed state. This is the signal of the transition from a quantum to a classical system. The kicks distributed by the bath can (with roughly the same probability of $40\%$ in equilibrium, i.e.$\!$ for $t\gtrsim 5$ fm$/c$) either bring about a feed-down mechanism of the excited state to the ground state, or melt the bound state. The final probability of remaing in the excited state is only about $20\%\,$.
\begin{itemize}
\item $l_{\psi_0} < l_{\rm env}< l_{\psi_1}$ regime
\end{itemize}
In this scenario the environment can resolve very well the excited state and also start affecting the ground state. From the diamond-shaped symbols on the left panels of Fig.\ref{fig:probab-entropy} one can see that an initial ground state has survival probability $P_0(t)\geq 80\%\,$, which is high but less than in the previous case. The chances of promoting the ground state to the excited state are slightly higher than before but still fewer than $10\%\,$. Similarly, the linear entropy grows more than before but stays below the value of $0.4\,$.
On the other hand, if the system is initially in the excited state (diamonds on the right panels of Fig.\ref{fig:probab-entropy}), the environment strongly perturbs the bound state, causing the linear entropy to increase abruptly and the excited state to disappear very rapidly, reaching soon the equilibrium value of about $10\%$ for the survival probability $P_1$. One might think that the disappearance of the excited state is simply a signal of its melting due to the kicks received from the bath. However, the feed-down mechanism is more important here than in the previous case, in which the excited state receives lighter kicks. In fact the probability of ending up with the ground state grows rapidly from zero to the equilibrium value of $60\%\,$, which leaves a melting probability of $30\%\,$. 
\begin{itemize}
\item $l_{\rm env}< l_{\psi_0}$ regime
\end{itemize}
Here the environment can fully resolve both bound states, causing a sudden increase of the linear entropy for any initial density matrix (see triangles in lower panels of Fig.\ref{fig:probab-entropy}). If the system is in the ground state at $t=0\,$, its survival probability will reduce quite quickly to the stationary value of about $55\%\,$. The ground state may then form an excited state with $P_1\approx 15\%$, or melt with a probability of approximately $30\%\,$. The case corresponding to the excited state at $t=0$ is basically identical to the previous scenario. The only difference is that the survival probability of the excited state is slightly higher than before, and the linear entropy reaches a lower equilibrium value of $0.5\,$.
Observe that, in this case, the equilibrium values for the probabilities and linear entropy for the two different initial density matrices are very similar. This is expected since in the limit $l_{\rm env}\to 0$ the momentum kicks coming from the bath have such a much larger value than the typical binding energies of the subsystem that, at long time scales, it does not matter which particular initial state the subsystem was prepared in.\\
In this paragraph only the initial conditions corrisponding to the ground and excited states have been considered. However, one can start off with an initial density matrix corresponding to any linear combination of ground and excited state. This is an important feature if one wants to consider a realistic experimental initial condition in which both the ground and excited states of a system are occupied.


\subsection{Role of scattering states}

\begin{figure}[t!]
\begin{center}
\includegraphics[width=8.7cm]{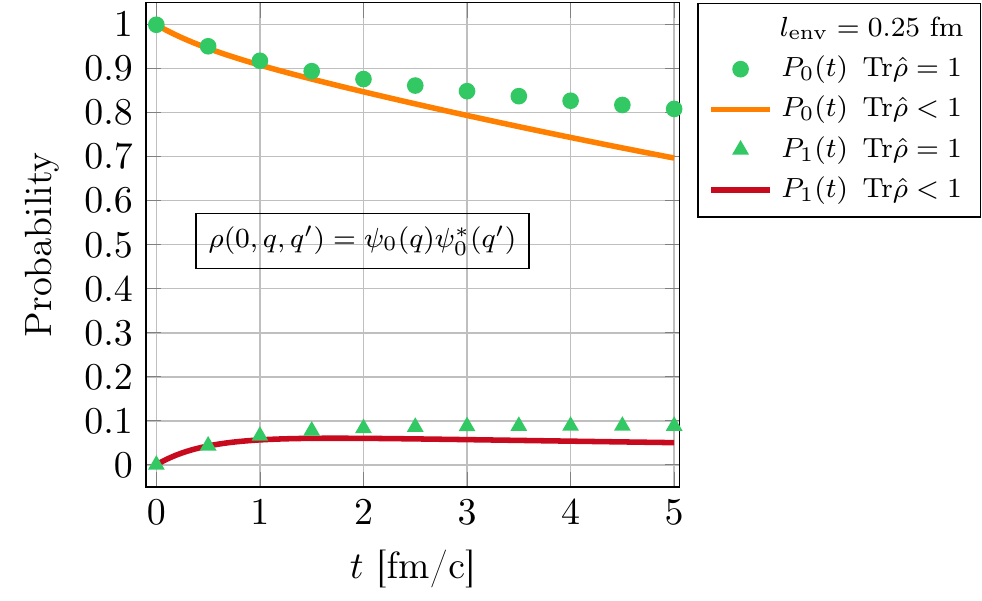}
\includegraphics[width=6.2cm]{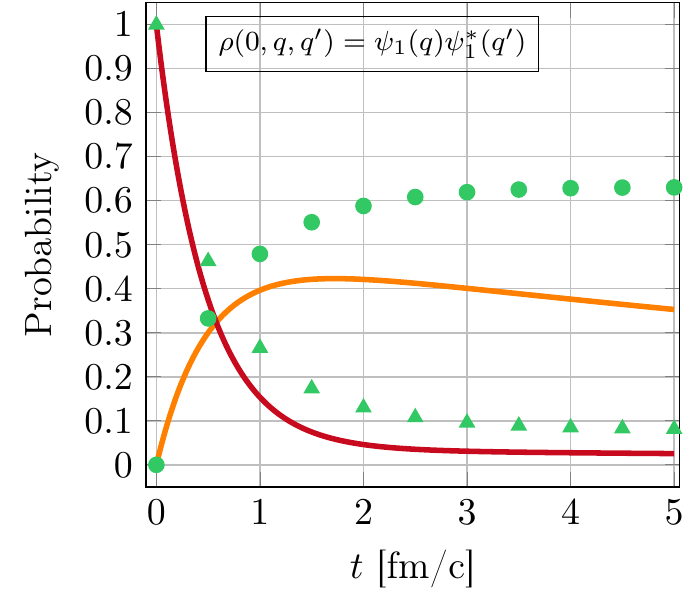}
\end{center}
\caption{Probabilities $P_0(t)\,,\,P_1(t)$ of having respectively the ground state $\psi_0$ and the excited state $\psi_1$ at time $t\,$, starting off with an initial density matrix $\rho_0(q,q')=\rho(0,q,q')$ corresponding to the ground state (on the left) or excited state (on the right). The physical case corresponds to the discrete data points, obtained by solving the Lindblad equation (\ref{eq:LindMath}), which preserves the trace of the density operator. The continuous lines represent the probabilities computed using the ansatz (\ref{eq:wrongexp}), which does not preserve probability.}
\label{fig:probabilities}
\end{figure}

It is clear that the presence of the scattering states in the Hilbert space of the subsystem is crucial for preserving probability, that is $\mbox{Tr}\hrho(t)=1\,$.
However, one might think that the presence of the scattering states is not that important in certain cases. After all, the previous paragraph has shown that, for certain correlation lengths of the medium, the probability of bound states dissociation or, equivalently, the formation of scattering states is small.
For example, the case $l_{\psi_0} < l_{\rm env}< l_{\psi_1}$ shows that $P_{\rm scatt}(t)=1-P_0(t)-P_1(t)\lesssim 0.1\,(0.3)\:\forall t$ when starting off with the ground (excited) state. Consequently one might hope that for this scenario the bound states probabilities $P_0(t)$ and $P_1(t)$ could be approximately computed without solving the full Lindblad equation for the density matrix, but expanding the density matrix in terms of the bound states only
\be\label{eq:wrongexp}
&&\rho(t,q,q')=\sum_{n,m=0,1}\!\!\!\lambda_{nm}(t)\psi_n(q)\psi_m^*(q')\:,\nn\\
&&\lambda_{nm}(t) = \int\!\!\diff{q}\!\!\int\!\!\diff{q'}\,\psi_n^*(q)\psi_m(q')\rho(t,q,q')\:,
\ee
and solving for the coefficient $\lambda_{nm}(t)$ by taking the time derivative
\be
\dot\lambda_{nm}(t) = \int\!\!\diff{q}\!\!\int\!\!\diff{q'}\,\psi_n^*(q)\psi_m(q')\pderiv{\rho(t,q,q')}{t}
\ee
and using the Lindblad equation (\ref{eq:LindMath}) with the replacement $q=r+y$ and $q'=r-y\,$.
One then solves the linear first-order differential equations for the probabilities $\lambda_{00}(t)=P_{0}(t)$ and $\lambda_{11}(t)=P_{1}(t)\,$, which are decoupled from the equations for the mixed coefficients $\lambda_{01}(t)$ and $\lambda_{10}(t)\,$.
Observe that the ansatz (\ref{eq:wrongexp}) does not take scattering states into account, hence does not preserve the trace of the density matrix. This technique has been used in the literature (e.g.$\!$ \cite{Akamatsu:2011se}) for computing approximate bound states survival probabilities.\\ 
Fig.\ref{fig:probabilities} shows that, if ansatz (\ref{eq:wrongexp}) is assumed, the time evolution of the bound states probabilities are quite different from the correct ones computed from the full Lindblad equation, which implicitly knows about the whole Hilbert space of the subsystem, which comprises the scattering states. Notice that the evolution of the probability of the excited state $P_1(t)$ with the ansatz (\ref{eq:wrongexp}) is similar to the correct one computed from the Lindblad equation. The situation is drastically different for the ground state, whose probability takes values significantly lower than the exact ones. This happens because (\ref{eq:wrongexp}) implies that all the states that ``leak out'' from the potential well have no chance of going back inside it, since the ansatz does not take into account scattering states.\\
The next subsection is going to show more explicitly that scattering states indeed roll back into the potential well to form the ground state with a non negligible probability. The feed-down mechanism from a scattering state to an excited state is instead a marginal effect, shedding light on why the excited-state probabilities calculated either via solving the full Lindblad equation or using the anstaz (\ref{eq:wrongexp}) are quite similar.


\subsection{Recombination}

\begin{figure}[t!]
\begin{center}
\includegraphics[width=7.5cm]{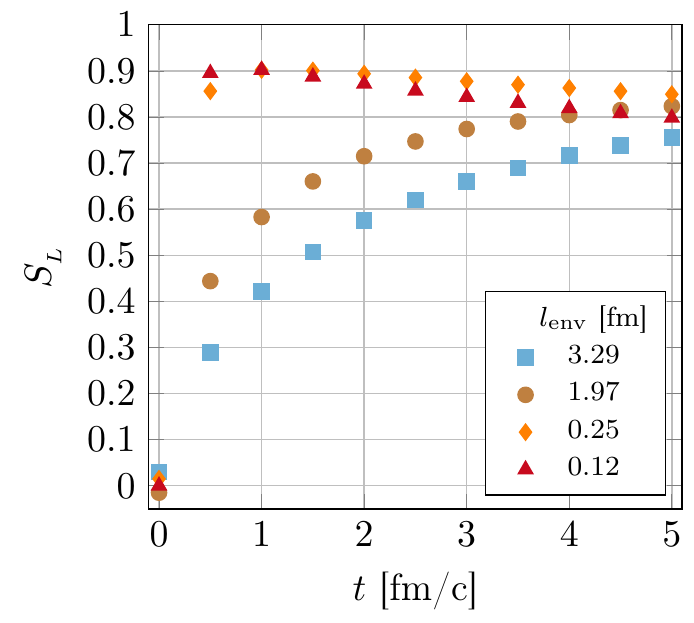}
\caption{Time evolution of the linear entropy for different correlation lengths of the environment, starting off with an initial density matrix corresponding to a thermal scattering state ($\,\delta=2.5$ fm and $p=-0.69$ GeV).}
\label{fig:scattentropy}
\end{center}
\end{figure}

\begin{figure}[t!]
\centering
\begin{minipage}[t]{0.48\textwidth}
\includegraphics[width=\textwidth]{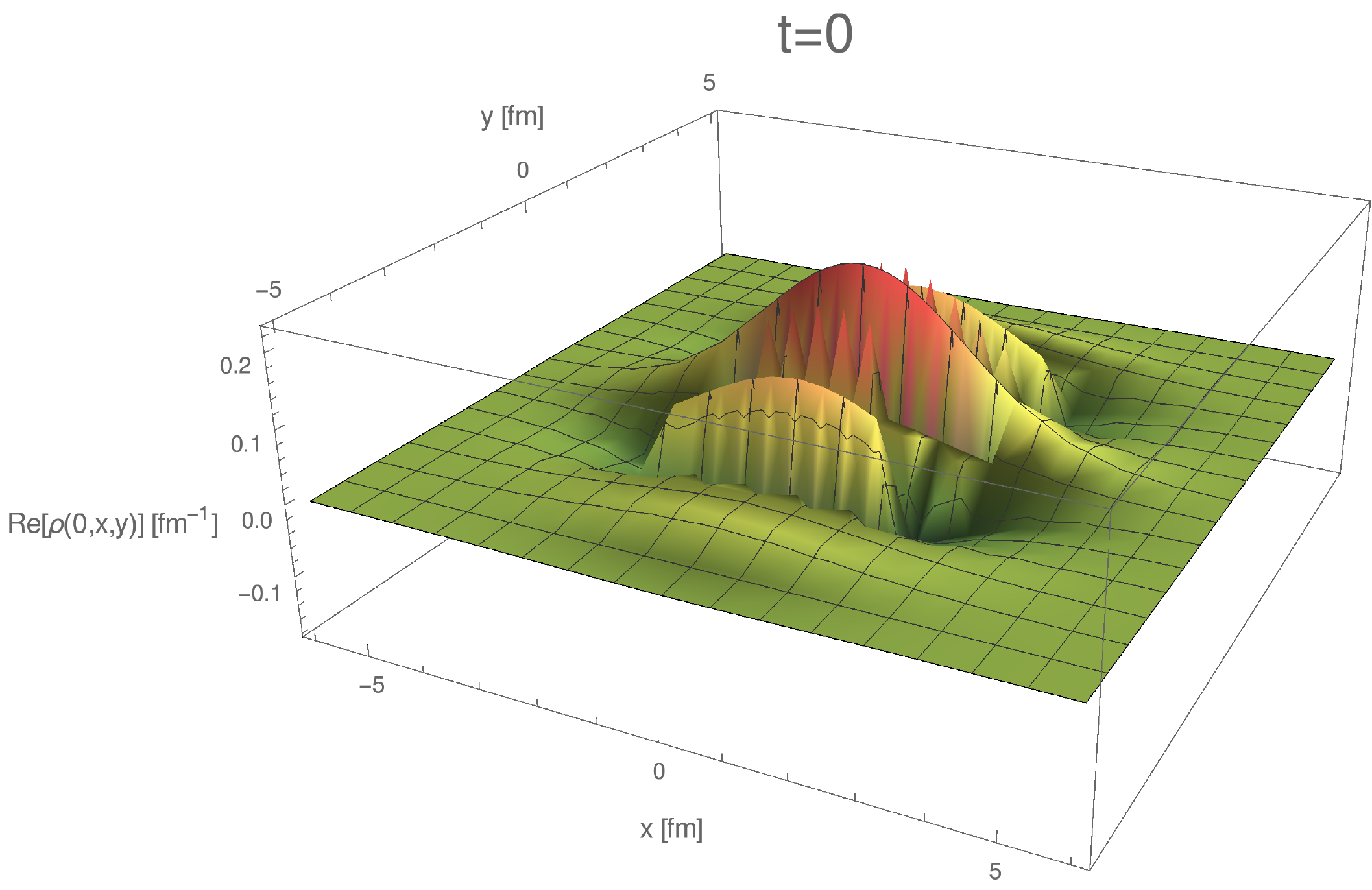}
\end{minipage}
\begin{minipage}[t]{0.48\textwidth}
\includegraphics[width=\textwidth]{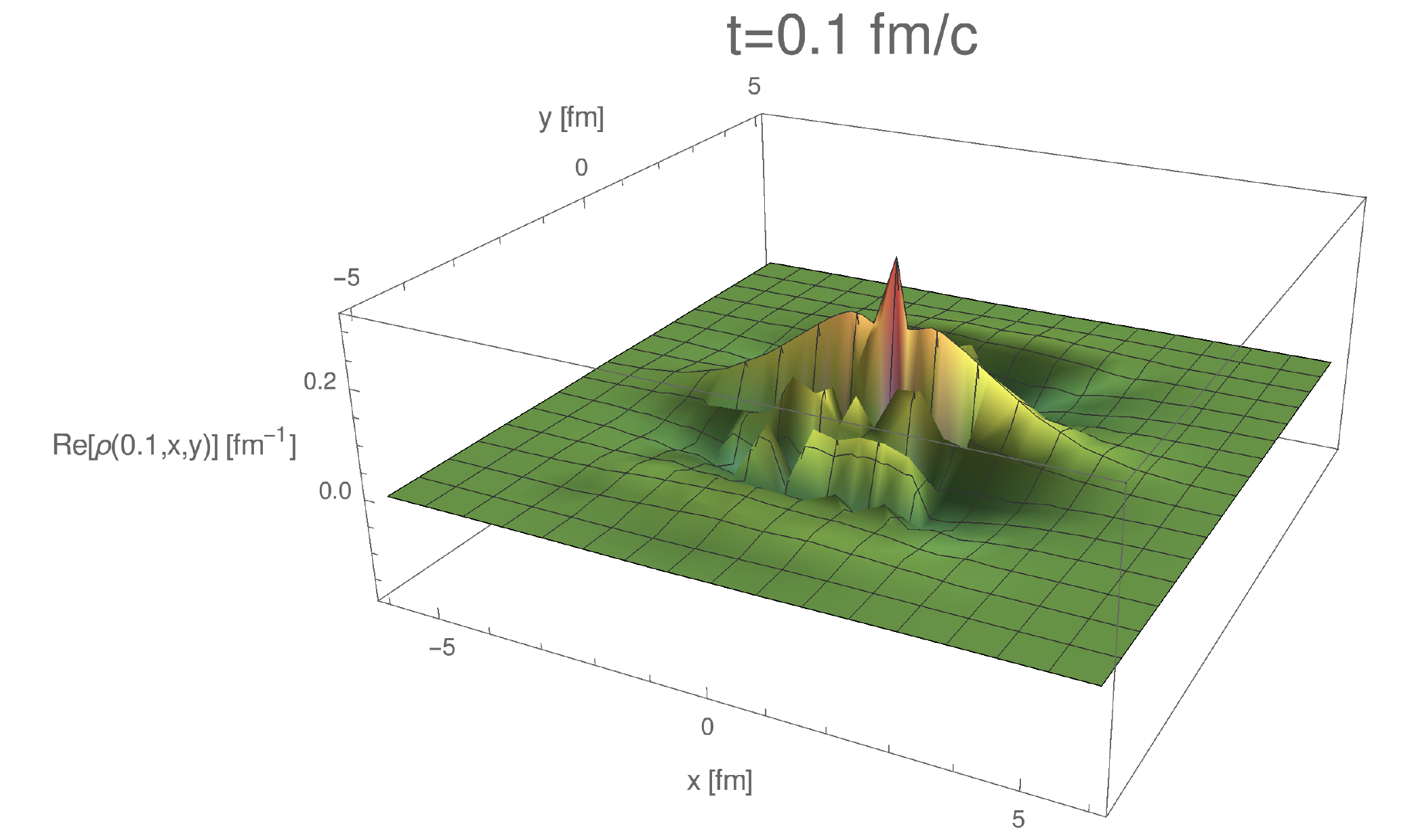}
\end{minipage}
\begin{minipage}[t]{0.48\textwidth}
\includegraphics[width=\textwidth]{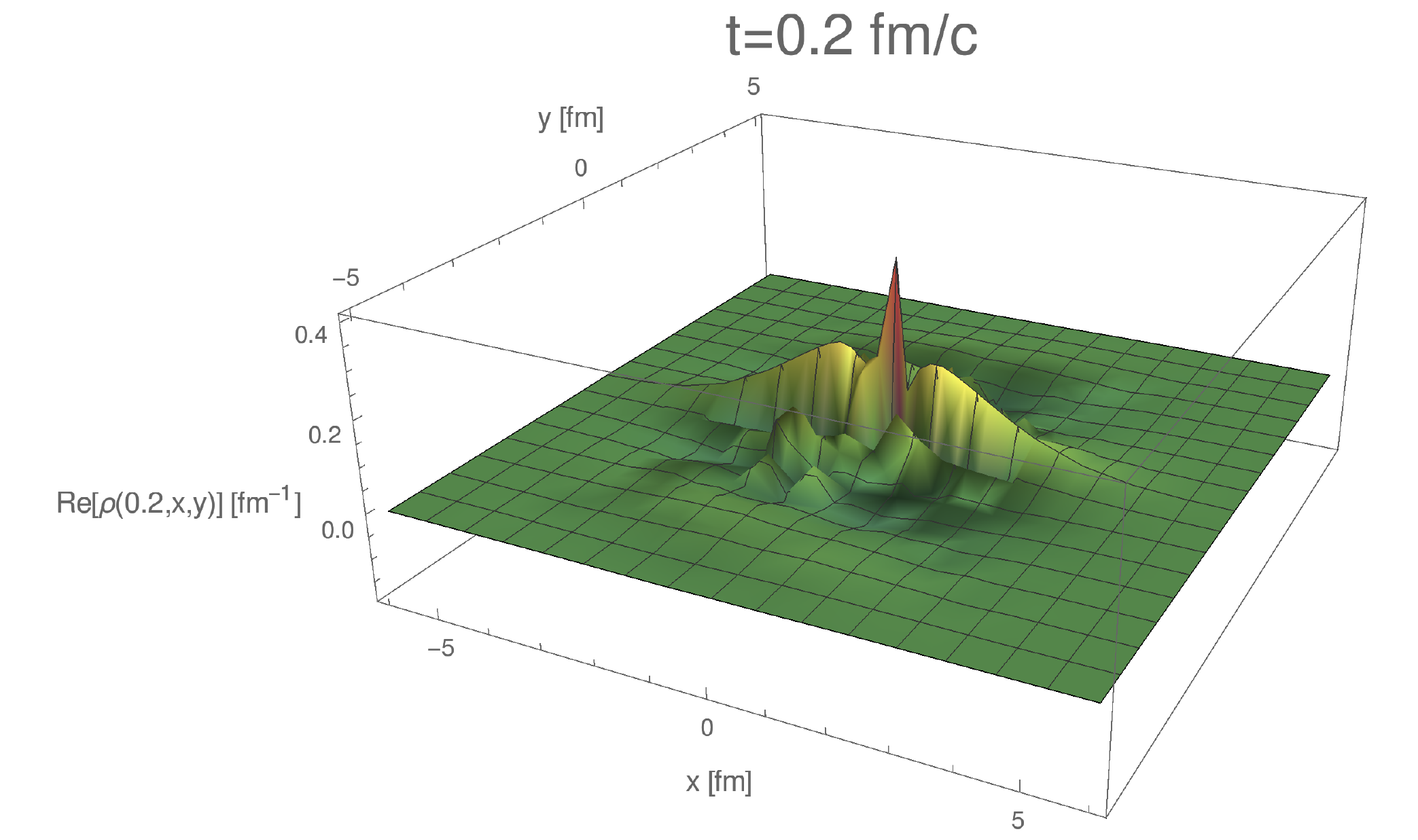}
\end{minipage}
\begin{minipage}[t]{0.48\textwidth}
\includegraphics[width=\textwidth]{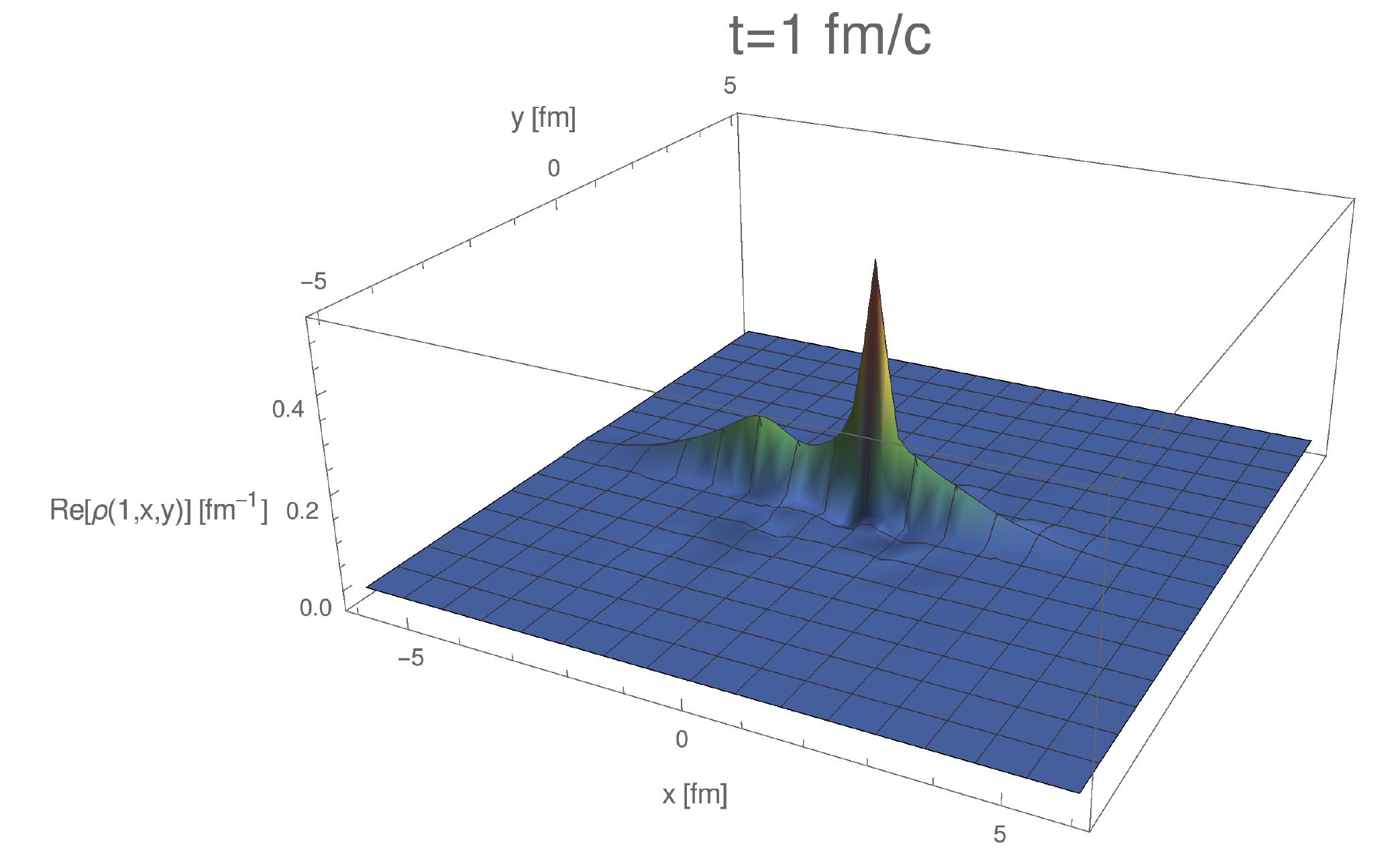}
\end{minipage}
\begin{minipage}[t]{0.48\textwidth}
\includegraphics[width=\textwidth]{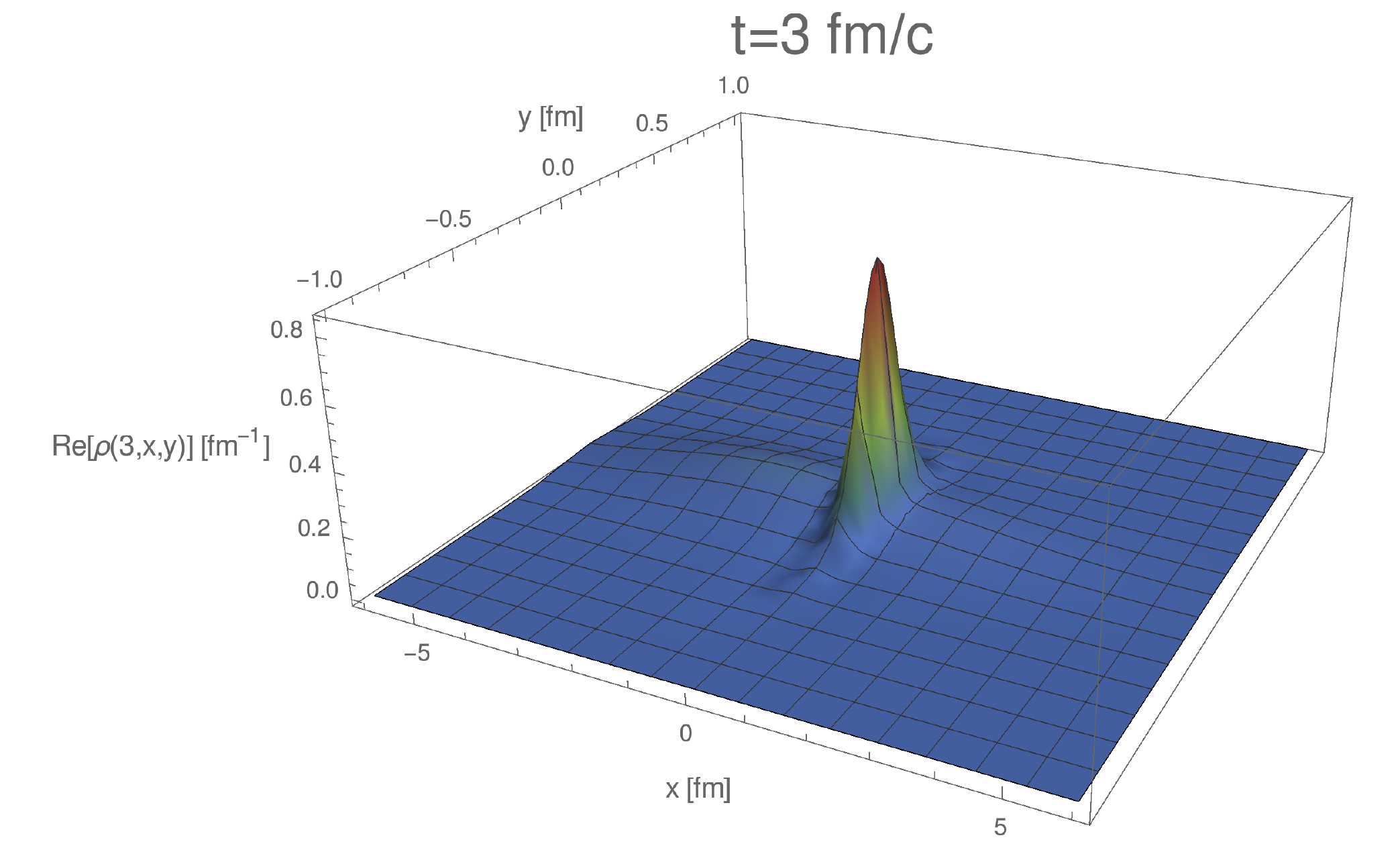}
\end{minipage}
\begin{minipage}[t]{0.48\textwidth}
\includegraphics[width=\textwidth]{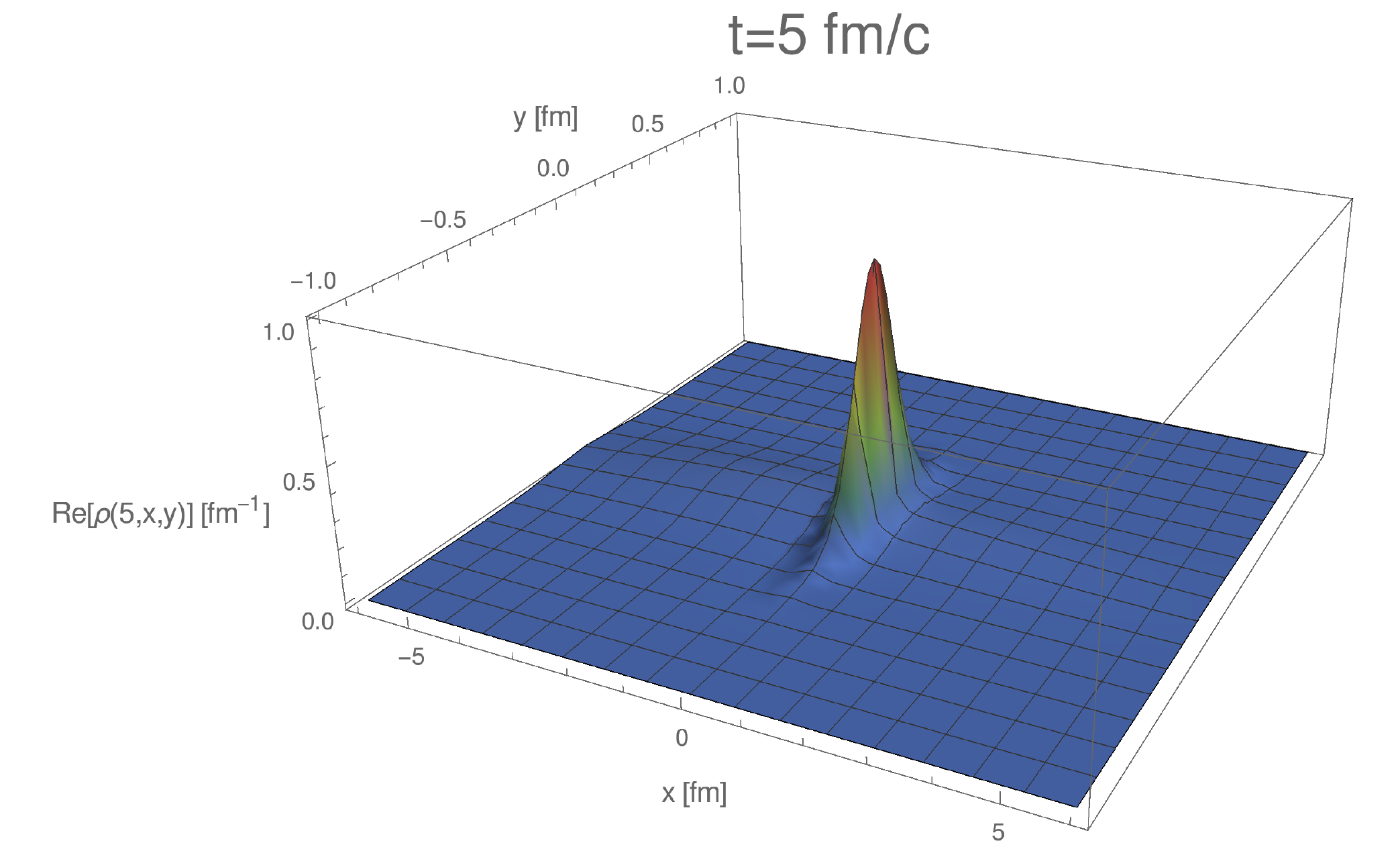}
\end{minipage}
\caption{Time evolution of the real part of the density matrix for the thermal scattering state defined before, propagating through a bath with $l_{\rm env}=0.25$ fm. Top: The medium disturbs the system at the very early times. Centre: Quantum decoherence appears as a suppression of the off-diagonal elements of the density matrix. Bottom: The state is squeezed around $y=0\,$. Notice the different scale in the $y$ direction.}
\label{fig:3dplots}
\end{figure}

\begin{figure}[htbp]
\centering
\begin{minipage}[t]{0.48\textwidth}
\includegraphics[width=\textwidth]{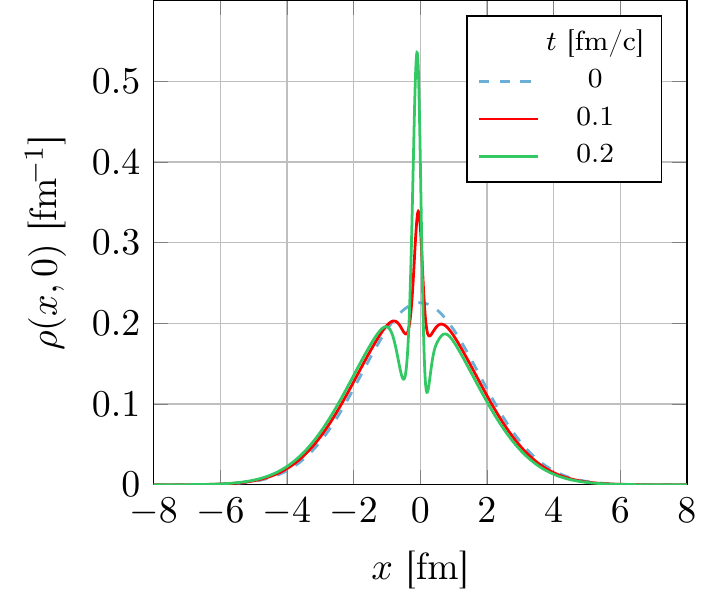}
\end{minipage}
\begin{minipage}[t]{0.48\textwidth}
\includegraphics[width=\textwidth]{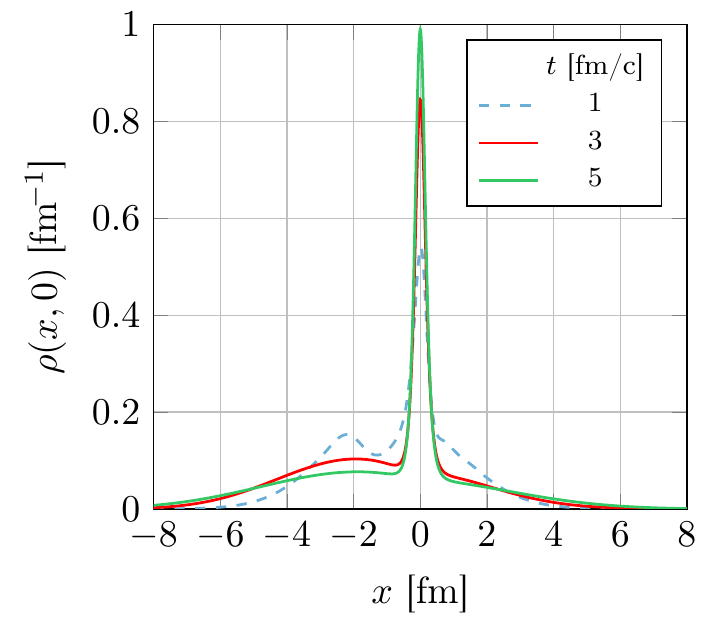}
\end{minipage}
\caption{Time evolution of the diagonal elements of the density matrix for the thermal scattering state defined before, propagating through a bath with $l_{\rm env}=0.25$ fm. It is clear that positivity is preserved by the Lindblad equation. The asymmetry of $\rho(t,x,0)$ for $t>0$ comes from the negative momentum of the state at $t=0\,$, which drives the expectation value of the position $\hat x$ towards negative values. Notice the different vertical scales in the two plots.}
\label{fig:positivity}
\end{figure}

This paragraph aims to show how (re)combination\footnote{The term recombination is used when the scattering state originally came from a melted $q\bar q$ bound state.} of a scattering state into a bound state takes place. In order to implement vanishing spatial boundary conditions for the density matrix when solving the Lindblad equation (\ref{eq:LindMath}), one has to consider normalised scattering states (with positive energies) defined as
\be\label{eq:scatteringstate}
\psi_{\rm scatt}(x) = \frac{1}{\pi^{\frac{1}{4}}\sqrt{\delta}}\eu^{-\frac{1}{2}\lbr\frac{x}{\delta}\rbr^2+\frac{\iu}{\hbar}xp}\:,
\ee
where $\delta=\sqrt{2}\sqrt{\avg{\hat{x}^2}}$ is the localisation parameter and $p=\avg{\hat p}$ is the expectation value of the momentum. In principle, the localised scattering state can be written as a superposition of elements of the Hilbert state of the $T=0$ Hamiltonian with the unbounded potential (\ref{eq:unbounded}), but this can not be done in practice since the full Hilbert space of the subsystem is unknown. An interesting case to analyse is the evolution of a thermal scattering state, that is a scattering state with size equal to the thermal de Broglie wavelength, i.e.$\!$ $\sqrt{\avg{\hat{x}^2}}=\lambda_{\rm th}=h/p_{\rm th}\,$, with thermal momentum $p_{\rm th}=\sqrt{\mu\,T}\,$, $\mu=m/2\,$. The temperature $T=0.8$ GeV has been used again in the numerical analysis. 
\begin{figure}[t!]
\centering
\begin{minipage}[t]{0.48\textwidth}
\includegraphics[width=\textwidth]{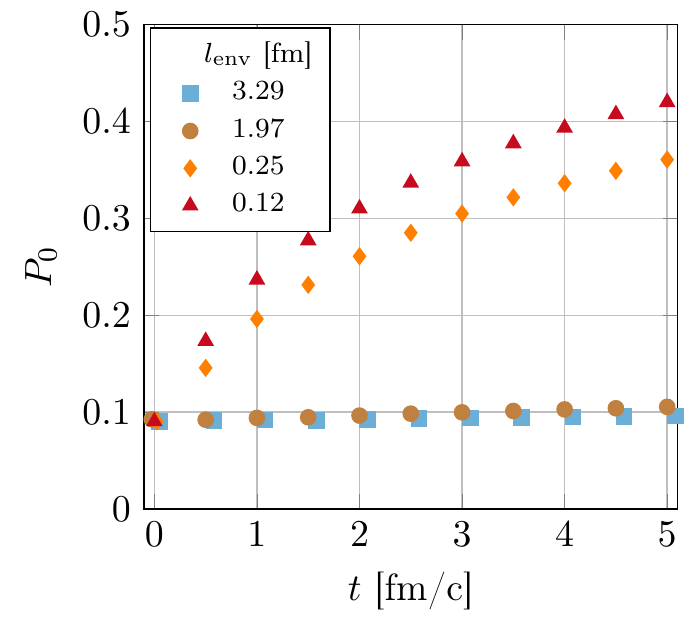}
\end{minipage}
\begin{minipage}[t]{0.48\textwidth}
\includegraphics[width=\textwidth]{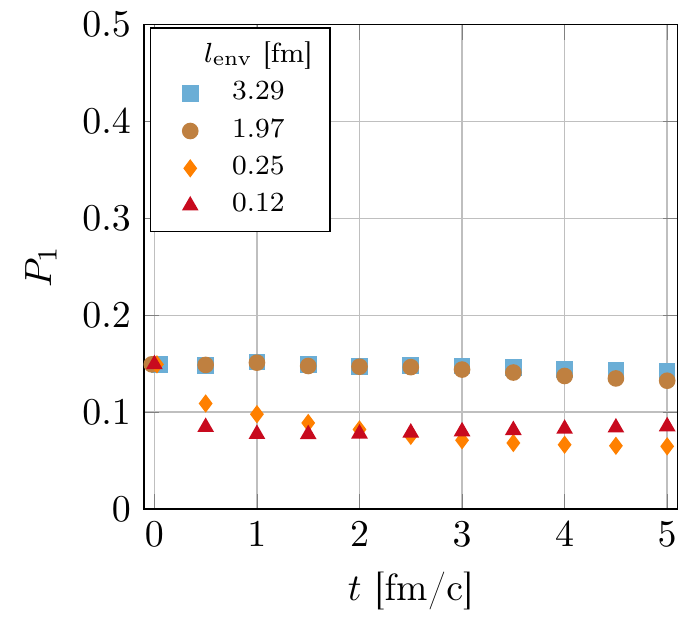}
\end{minipage}
\caption{On the left (right): Probability of having the ground (excited) state at time $t$ for different correlation lengths of the environment, starting off with an initial density matrix corresponding to a thermal scattering state ($\,\delta=2.5$ fm and $p=-0.69$ GeV). Recall that the typical sizes of the quantum states are $\sqrt{\avg{x^2}}_{\psi_0}=0.162$ fm, $\sqrt{\avg{x^2}}_{\psi_1}=0.384$ fm and $\sqrt{\avg{x^2}}_{\psi_{_{\rm scatt}}}=\delta/\sqrt{2}=1.77$ fm. Some of the points are slightly shifted horizontally to avoid superpositions of symbols.}
\label{fig:scattprobabilities}
\end{figure}
Fig.\ref{fig:scattprobabilities} shows the probability of the thermal scattering state to become a bound state (either the ground or excited state) for different $l_{\rm env}\,$. When $l_{\rm env}>\sqrt{\avg{x^2}}_{\psi_{_{\rm scatt}}}\,$, both $P_0$ and $P_1$ do not vary with time (see square-shaped symbols). This entails that the recombination probability, defined as $P_{\rm rec}(t)=\sum_{i=0,1}\lbr P_i(t)-P_i(0)\rbr$, remains practically zero. A very similar behaviour is found for $l_{\rm env}\sim\sqrt{\avg{x^2}}_{\psi_{_{\rm scatt}}}$ (see circle-shaped symbols), with the only difference that here the probability of getting the ground state slightly increases with time, whereas the probability of getting the excited state slightly diminishes.\\
For these two correlation lengths of the environment the linear entropy grows relatively rapidly (see Fig.\ref{fig:scattentropy}). The scattering state is indeed a superposition of eigenstates of the Hamiltonian of the subsystem with the modified potential (\ref{eq:unbounded}), and the medium disturbs all the eigenstates with wavelengths either bigger or of the same order of $l_{\rm env}$.
This effect becomes obviously stronger when $l_{\rm env}$ becomes smaller than the size of the scattering state, causing the linear entropy to raise abruptly up to values very close to the maximum value of one. This means that the subsystem has been strongly perturbed by the medium and the quantum information is quickly lost. The fact that the dynamics becomes basically classical does not prevent the subsystem to form bound states, as can be seen from the finite recombination probability inferred from Fig.\ref{fig:scattprobabilities} (see diamonds and triangles). In fact the probability of the scattering state to form the ground state, in the case with $l_{\rm env}=0.12$ fm, grows from the $t=0$ overlap value of approximately $9\%$ to $42\%$ at $t=5$ fm/c. On the other hand, the overlap with the excited states decreases of about $6\%\,$. Observe that the long-time value of the linear entropy for the two smallest values of $l_{\rm env}$ is determined very accurately by $P_0(t)$ only. For example, for $l_{\rm env}=0.12$ fm one has $0.8= S_{_L}\approx 1-P_0^2= 0.82$ at $t=5$ fm/c.\\
$\hspace{-0.34cm}$Eventually Fig.\ref{fig:3dplots} showcases some direct results obtained by solving the Lindblad equation (\ref{eq:LindMath}) with $l_{\rm env}=0.25$ fm, starting off with the thermal scattering state defined above. It is clear that quantum decoherence manifests itself at the very early stages via the suppression of the off-diagonal elements ($y\neq 0$) of the real part of the density matrix. After barely $\Delta t=0.2$ fm/c the density matrix considerably changes (explaining the sudden raise of the linear entropy in Fig.\ref{fig:scattentropy}), and at $t=1$ fm/c approaches a diagonal form. The imaginary part of $\rho$ has not been displayed since it goes close to zero very quickly, leaving the density matrix basically real.\\
By considering just the time evolution of the diagonal elements of the density matrix, one can check that positivity is indeed preserved, that is $\rho(t,x,0)\geq 0$ for all positions $x$ and times $t$ ($\,\rho(t,x,0)\in \mathbb{R}$ $\forall x,t\,$), as can be seen in Fig.\ref{fig:positivity}\footnote{Note that positivity of the density operator comes automatically from the Lindblad form (\ref{eq:Lindblad}). However, equation (\ref{eq:LindMath}) is an \textit{almost} exact Lindblad equation, due to the very small terms neglected in perturbation theory. Therefore it can happen that in certain regimes, in which the approximations used throughout this work are not valid anymore, the density operator sligthly violates positivity.}.

$\hspace{-0.75cm}$In order to see how the dissociation and recombination mechanisms depend on the mass of the heavy quarks, one can find the numerical results for the bound-state probabilities for a heavier quark with mass ($m=4.7$ GeV) in Appendix \ref{masses}.

\section{Summary and Outlook}\label{sec:end}

It has been shown how to derive a Lindblad equation for non relativistic heavy quarks and antiquarks propagating out of equilibrium in a thermalised quark gluon plasma, within the framework of open quantum systems and starting from the underlying gauge theory. To achieve this, an abelian model
for the plasma has been used, together with some well-defined approximations, including the perturbative expansion and the Markovian limit.
All the Lindbladian terms in the master equation depend on the imaginary part of the inter-quark potential, whose shape gives the correlation length of the environment.
Different types of dynamics of the heavy probes emerge according to the value of the ratio between the correlation length of the environment and the typical size of the probes. The smaller the correlation length of the medium, the higher the resolution of the quantum probes, giving rise to quantum decoherence and to classical dynamics.
Numerical simulations of the one-dimensional Lindblad equation for a heavy quark-antiquark pair allow one to study quantitatively the mechanisms of melting and formation of bound states within a unique framework. It has been found that the interaction of bound states with the medium can make the bound state dissociate or undergo a feed-down mechanism to a bound state with lower energy. Similarly, a scattering state (that can be interpreted as a melted bound state) has a non-zero probability of (re)combining into a bound state.\\
In principle, dissociation and formation of bound states can also be analysed starting from the same initial density matrix by solving the Lindblad equation for more than one heavy $q\bar q$ pair, which can be straightforwardly derived with the same technique exploited here for the two-particle case.
The issue with this procedure is that a numerical simulation of such equation becomes very challenging for more than two particles, already in one dimension.
A solution to this problem would be to solve the stochastic Schr\"odinger equation \cite{PhysRevD.13.857,0305-4470-25-21-023,Gisin1989} associated to the Lindblad equation. The advantage of this approach is that the dimension of a wave vector is the square root of that of a density matrix. This strategy has been exploited by \cite{Akamatsu:2011se} in one dimension in the framework of quarkonium in a quark gluon plasma, but only the recoilless limit (no friction) has been taken into account.\\
Another interesting and numerically feasible approach to study the real-time dynamics of many particles in the plasma is the Langevin equation, which has been proven to be the correct description when the system has undergone a quantum to classical transition. However, in order to be able to use the classical stochastic equation, one needs to understand how to implement the initial quantum state correctly, since the Langevin formalism is strictly valid only after a time scale of the order of the decoherence time.
It has been found that the Langevin equation derived from the gauge theory for a $q\bar q$ pair (see also \cite{Blaizot:2015hya} for more particles) features a space-dependent friction coefficient, which encodes physical information about polarisation effects of the medium induced by the relative position of the heavy probes. Apart from \cite{Blaizot:2015hya}, this feature has not yet been considered in the literature when solving the Langevin equation for quarkonia. It would be interesting to see whether this space-dependent friction for a heavy $q\bar q$ pair could be reproduced for strongly coupled plasmas in the context of the AdS/CFT correspondence, as it has already been done for a single heavy quark \cite{PhysRevD.74.085012,Son:2009vu}\footnote{This idea has been suggested by Prem Kumar. Observe that single-quark diffusion and heavy $q\bar q$ pairs are treated differently in holography.}.\\
Eventually, in order to envisage some phenomenological applications, a natural development of this project would be to obtain a Lindblad equation in QCD using the same formalism presented here (see \cite{Akamatsu:2012vt,Akamatsu:2014qsa,Brambilla:2016wgg} for other approaches). One would expect to be able to implement non abelian features within the present scheme, this time considering the weak coupling approximation more carefully, being the non-linear QCD couplings less strongly suppressed than the QED ones.
This route has been followed in the preliminary work \cite{prep}, in which the QED case can be obtained as a restriction of the QCD case.


\section*{Acknowledgments}

I am very grateful to Jean-Paul Blaizot and Pietro Faccioli for fruitful discussions on theoretical aspects of this paper and for having invited me at IPhT (Saclay) and Trento University respectively.
A special acknowledgment goes to Giovanni Garberoglio, who carefully checked all the numerical results, and to Gert Aarts, who provided interesting comments and did a meticulous job in proofreading the manuscript.
I am also thankful to Prem Kumar, Jamie Mc Donald, Guy Jehu and Markus Muller for valuable remarks and suggestions,
and to my supervisor Chris Allton, for giving me the freedom to pursue this project.


\appendix

\section{Derivation of the Lindblad equation for one heavy quark}\label{A3}

In order to obtain the Lindblad equation (\ref{eq:Lindblad1Q}) for one heavy quark one has to compute the matrix elements in (\ref{eq:Lindblad2}). The first one on the right hand side reads ($\int_\p\equiv\int\frac{\diff{\p}}{(2\pi\hbar)^3}$)
\be
-\bra{\q}\hH_{\rm eff}\hrho\ket{\q'}&=&-\int_{\x}\bra{\q}\hH_{\rm eff}\ket{\x}\rho(t,\x,\q')\nn \\
&=& -\frac{1}{2m}\int_{\x}\bra{\q}\hat\p^2\ket{\x}\rho(t,\x,\q') +\frac{\iu}{2}\sum_\mu\int_{\x}\bra{\q}\hLmu^\da\hLmu\ket{\x}\rho(t,\x,\q')\nn\\
&\stackrel{(\ref{eq:Wigner-1})}{=}& - \frac{1}{2m}\int_{\x}\rho(t,\x,\q')\int_\p\p^2\eu^{\frac{\iu}{\hbar}(\q-\x)\cdot\p}\nn\\
&& +\frac{\iu}{2}\sum_\mu\int_{\x}\rho(t,\x,\q')\int_\p\left|\Lambda_\mu\lbr \frac{\q+\x}{2},\p \rbr\right|^2\eu^{\frac{\iu}{\hbar}(\q-\x)\cdot\p}\nn\\
&=& \lbr\frac{\hbar^2}{2m}\pderiv[2]{}{\q} + \frac{\iu}{2}\sum_\mu\left|\Lambda_\mu\lbr \q\rbr\right|^2 \rbr \rho(t,\q,\q')\:,
\ee
where it has been used that $|\Lambda_\mu|^2$ is momentum independent up to order $g^2$. The same calculation applies to the second term on the right hand side of (\ref{eq:Lindblad2}) and gives expression (\ref{eq:2}) as a result.
The computation of the last term in (\ref{eq:Lindblad2}) reads
\be
&&\hspace{-1cm}-\iu\sum_\mu\bra{\q}\hLmu\hrho\hLmu^\da\ket{\q'} \nn\\
&&\hspace{-1cm}=-\iu\sum_\mu\int_{\x}\int_{\y}\,\rho(t,2\x-\q,2\y-\q')\int_\p\int_\l\Lmu\lbr x,\frac{p}{2} \rbr\Lmu^*\lbr y,\frac{l}{2} \rbr\eu^{\frac{\iu}{\hbar}\p\cdot(\q-\x)}\eu^{-\frac{\iu}{\hbar}\l\cdot(\q'-\y)}\nn\\
&\stackrel{(\ref{eq:LL*})}{=}& -\iu\int_{\x}\int_{\y}\,\rho(t,2\x-\q,2\y-\q')\int_\p\int_\l\eu^{\frac{\iu}{\hbar}\p\cdot(\q-\x)-\frac{\iu}{\hbar}\l\cdot(\q'-\y)}\nn\\
&&\lbr\iu\varphi_{\tau}\lc\x,\frac{\p}{2};\y,\frac{\l}{2}\rc-\frac{1}{2}D^2_{\tau}[\x;\y] + \frac{1}{2}\sum_\mu\lc|\Lambda_\mu(\x)|^2 + |\Lambda_\mu(\y)|^2\rc\rbr
\nn\\
&&\hspace{-1cm}=\int_{\x}\int_{\y}\,\rho(t,2\x-\q,2\y-\q')\int_\p\int_\l\varphi_{\tau}\lc\x,\frac{\p}{2};\y,\frac{\l}{2}\rc\eu^{\frac{\iu}{\hbar}\p\cdot(\q-\x)-\frac{\iu}{\hbar}\l\cdot(\q'-\y)}\nn\\
&&\hspace{-1cm}+\frac{\iu}{2}\lbr D_{\tau}^2[\q;\q']- \sum_\mu\lc|\Lambda_\mu(\q)|^2 + |\Lambda_\mu(\q')|^2\rc \rbr \rho(t,\q,\q')\:.
\ee
Collecting all the results as shown in section \ref{sec:1Q}, one finally gets to the Lindblad equation (\ref{eq:Lindblad1Q}).


\section{Derivation of the Ehrenfest equations of motion}\label{A2}

Here the proof of the first two equations of motion in
(\ref{Ehrenfest}) is shown. The derivation of the other relations is very similar and is therefore omitted here.
The Ehrenfest equation of motion for the first momentum of the position operator is obtained through the following steps (hats on operators have been removed for ease of notation): 
\be
&&\!\!\!\!\!\deriv{}{t}\avg{q}=\mbox{Tr}\left[q\,\deriv{}{t}\rho(t)\right]\nn\\
&&\!\!\!\!\!\stackrel{(\ref{eq:Caldeira2})}{=}-\frac{\iu}{\hbar}\mbox{Tr}\left[q\left[ H,\rho(t)\right]\right]-\frac{\iu\gamma}{2\hbar}\,\mbox{Tr}\left[q\left[q,\left\{p,\rho(t)\right\}\right]\right]-\frac{\kappa}{2\hbar^2}\,\mbox{Tr}\left[q\left[q,\left[q,\rho(t)\right]\right]\right]\nn\\
&&\!\!\!\!\!=-\frac{\iu}{\hbar}\mbox{Tr}\left[q\left[ H,\rho(t)\right]\right]=-\frac{\iu}{2m\hbar}\mbox{Tr}\left[q\left[p^2,\rho(t)\right]\right]\:,
\ee
where in the last line the cyclicity of the trace has been exploited to cancel
the friction and diffusion terms. The fact that the position
operator commutes with the potential operator has been also used. The commutator
$[q,p]=\iu\hbar$ then gives
\be
\deriv{}{t}\avg{q}&=&-\frac{\iu}{2m\hbar}\left(\iu\hbar\,\mbox{Tr}\left[p\,\rho(t)\right] +\mbox{Tr}\left[pqp\,\rho(t)\right]-\mbox{Tr}\left[q\,\rho(t)pp\right]\right)\nn\\
&=&-\frac{\iu}{2m\hbar}\left(2\iu\hbar\,\mbox{Tr}\left[p\,\rho(t)\right] +\mbox{Tr}\left[ppq\,\rho(t)\right]-\mbox{Tr}\left[q\,\rho(t)pp\right]\right)\nn\\
&=&\frac{1}{m}\,\mbox{Tr}\left[p\,\rho(t)\right]=\frac{\avg{p}}{m}\:.
\ee
The second relation of (\ref{Ehrenfest}) follows from a similar
calculation. One has
\be\label{firstcontrib}
&&\!\!\!\!\!\deriv{}{t}\avg{p}=\mbox{Tr}\left[p\,\deriv{}{t}\rho(t)\right]\nn\\
&&\!\!\!\!\!\stackrel{(\ref{eq:Caldeira2})}{=}-\frac{\iu}{\hbar}\mbox{Tr}\left[p\left[ H,\rho(t)\right]\right]-\frac{\iu\gamma}{2\hbar}\,\mbox{Tr}\left[p\left[q,\left\{p,\rho(t)\right\}\right]\right]-\frac{\kappa}{2\hbar^2}\,\mbox{Tr}\left[p\left[q,\left[q,\rho(t)\right]\right]\right]\nn\\
&&\!\!\!\!\!=-\frac{\iu}{\hbar}\mbox{Tr}\left[p\left[ H,\rho(t)\right]\right]-\frac{\iu\gamma}{2\hbar}\,\mbox{Tr}\left[p\left[q,\left\{p,\rho(t)\right\}\right]\right]\:,
\ee
since
\be\label{0}
\mbox{Tr}\left[p\left[q,\left[q,\rho(t)\right]\right]\right]=\mbox{Tr}\left[[p,q]\left[q,\rho(t)\right]\right]=
-\iu\hbar\mbox{Tr}\left[\left[q,\rho(t)\right]\right] =0
\ee
The first contribution in (\ref{firstcontrib}) becomes
\be\label{1}
-\frac{\iu}{\hbar}\mbox{Tr}\left[p\left[ H,\rho(t)\right]\right]&=&-\frac{\iu}{\hbar}\mbox{Tr}\left[p\left[ V_{\rm ext}(q),\rho(t)\right]\right]\nn\\
&=&-\frac{\iu}{\hbar}\mbox{Tr}\left[\left[p,V_{\rm ext}(q)\right]\rho(t)\right]-\mbox{Tr}\left[V'_{\rm ext}(q)\rho(t)\right]\nn\\
&=&-\avg{V'_{\rm ext}(q)}\:,
\ee
where the relation $\left[p,V_{\rm ext}(q)\right]=-\iu\hbar V'_{\rm ext}(q)$ has been used.
The second term of (\ref{firstcontrib}) reads
\be\label{2}
&&-\frac{\iu\gamma}{2\hbar}\,\mbox{Tr}\left[p\left[q,\left\{p,\rho(t)\right\}\right]\right]=-\frac{\iu\gamma}{2\hbar}\,\mbox{Tr}\left[\left[p,q\right]\left\{p,\rho(t)\right\}\right]\nn\\
&&=-\frac{\gamma}{2}\,\mbox{Tr}\left[\left\{p,\rho(t)\right\}\right]=-\gamma\,\mbox{Tr}\left[p\,\rho(t)\right]\nn\\
&&=-\gamma\,\avg{p}\:.
\ee
Expressions (\ref{firstcontrib}), (\ref{1}) and (\ref{2}) together give the second Ehrenfest relation
(\ref{Ehrenfest}). 


\section{From the Langevin equation with multiplicative noise to the Fokker-Planck equation}\label{A1}

A stochastic differential (Langevin) equation for a multi-component process can be written in the general form
\be\label{app:lang}
\deriv{S_i}{t}=A_i(\Sg,t)+\sum_jB_{ij}(\Sg,t)\eta_j(t)\:,
\ee 
with the white noises satisfying
\be
\avg{\eta_j(t)}_{\bm{\eta}}=0\:,\qquad \avg{\eta_j(t)\eta_k(t')}_{\bm{\eta}}=2\lambda\delta_{jk}\delta(t-t') \qquad \forall j,k\:.
\ee
In \cite{2007cond.mat..1242G} (see Sec. $\!$5.3.1) it is shown that the Fokker-Planck equation associated to the Markov process described by (\ref{app:lang}) is
\be\label{app:multi}
&&\pderiv{P(\Sg,t)}{t} = -\sum_i\pderiv{}{S_i}\lc\lbr A_i(\Sg,t)+\lambda\sum_{j,k}B_{jk}(\Sg,t)\pderiv{B_{ik}(\Sg,t)}{S_j}\rbr P(\Sg,t)\rc\nn\\
&&+\lambda\sum_{i,j}\frac{\partial^2}{\partial S_i\partial S_j}\lc \lbr \sum_{j,k}B_{ik}(\Sg,t)B_{jk}(\Sg,t) \rbr P(\Sg,t)\rc\:.
\ee
From now on $\lambda$ is set to $1/2\,$. The goal of this section is to prove that the Langevin equation (\ref{eq:Langmulti}) with multiplicative noise generates the Fokker-Planck-like equation (\ref{eq:Fokkermulti}). The Langevin equation (\ref{eq:Langmulti}) can be cast (up to o($g^2$) terms) in the multivariate form (\ref{app:lang}) as
\be\label{app:lang2}
&&\dot\r = \v -\frac{g^2\hbar\beta}{4\mu}\partial_\r W(\r)\:,\nn\\
&&\dot\v = -\frac{g^2\hbar\beta}{4\mu}\h(0)\v -\frac{g^2}{\mu}\partial_\r V(\r)+\frac{1}{\mu}G(\r)\eta(t)\:,
\ee
with the usual white noise $\eta(t)$ and the function $G(\r)$ defined via
\be\label{app:defD}
G^2(\r)\equiv\frac{g^2\hbar}{2}\lbr \h(0)+\h(\r)\rbr\stackrel{(\ref{eq:difflang})}{=}\kappa(\r)\:.
\ee
It is interesting to see that the first line of eq.(\ref{app:lang2}) does not correspond to the classical $\dot\r=\v\,$, but it contains an extra perturbative quantum correction that can be traced back to the modified classical equations of motion (\ref{eq:eqmotion}) derived from minimisation of the action containing the Lindbladian terms. In order to avoid a cumbersome notation, only the one-dimensional case is considered here.
The proof in $d>1$ dimensions works in the same way. Contrasting eq.(\ref{app:lang2}) with eq.(\ref{app:lang}) and using $\Sg=(S_1,S_2)\equiv(r,v)\,$, one obtains the following coefficients for the multivariate Langevin equation:
\be
&&A_1(r,v) = v - -\frac{g^2\hbar\beta}{4\mu}W'(r)\:;\qquad A_2(r,v) = -\frac{g^2\hbar\beta}{4\mu}W''(0)v -\frac{g^2}{\mu}V'(r)\:,\nn\\
&&B_{11}=B_{12}=B_{21}=0\:;\hspace{2.1cm} B_{22}(r)=\frac{G(r)}{\mu}\:,\nn\\
&&\eta_1=0\:;\hspace{4.8cm}\eta_2(t)=\eta(t)\:.
\ee
Substituting these terms in (\ref{app:multi}), one gets
\be
&&\!\!\!\!\pderiv{P(r,v,t)}{t} = -\pderiv{}{r}\lc \lbr v -\frac{g^2\hbar\beta}{4\mu}W'(r) \rbr P(r,v,t) \rc\nn\\
&&\!\!\!\!-\pderiv{}{v}\lc \lbr -\frac{g^2\hbar\beta}{4\mu}W''(0)v -\frac{g^2}{\mu}V'(r) \rbr P(r,v,t) \rc +\frac{1}{2}\pderiv[2]{}{v}\lc \frac{G(r)^2}{\mu^2} P(r,v,t) \rc\:,
\ee
whence, using (\ref{app:defD}) and $v=p/\mu\,$,
\be
&&\!\!\!\!\lc \partial_t + \lbr \frac{2p}{\mu}- \frac{g^2\hbar\beta}{4\mu}W'(r)\rbr -g^2V'(r)\partial_p\rc P(r,p,t) = \\
&&\!\!\!\!\frac{g^2\hbar}{4}\lbr W''(0)+W''(r)\rbr\partial_p^2 P(r,p,t) + \frac{g^2\hbar\beta}{4\mu}\lbr W''(0)+W''(r)+pW''(0)\partial_p \rbr P(r,p,t)\:,\nn
\ee
which has exactly the same form of the Fokker-Planck-like equation (\ref{eq:Fokkermulti}) for the Wigner function with $m=2\mu\,$.


\section{Mass dependence of dissociation and recombination}\label{masses}

\begin{figure}[t!]
\begin{center}
\includegraphics[width=8.5cm]{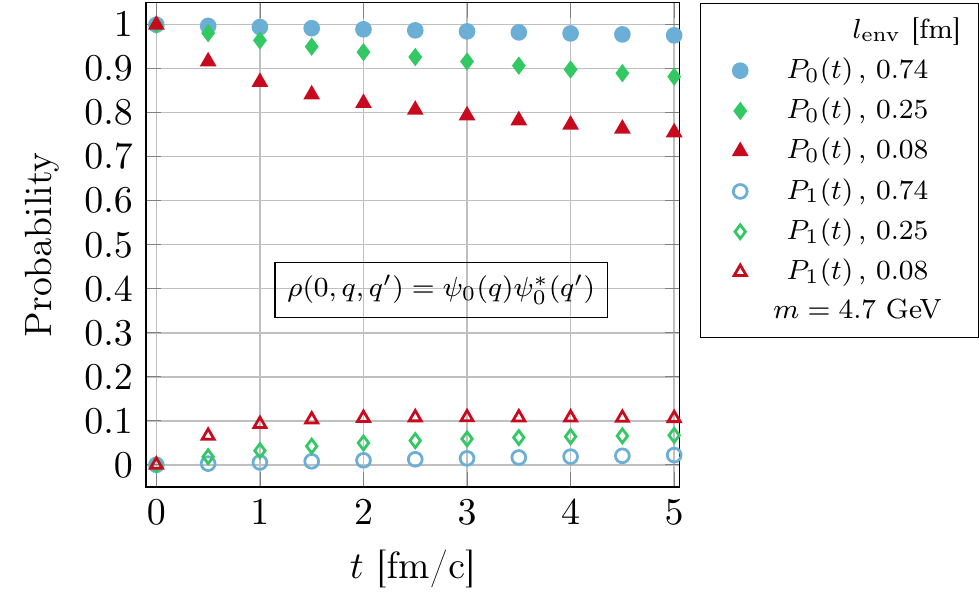}
\includegraphics[width=6cm]{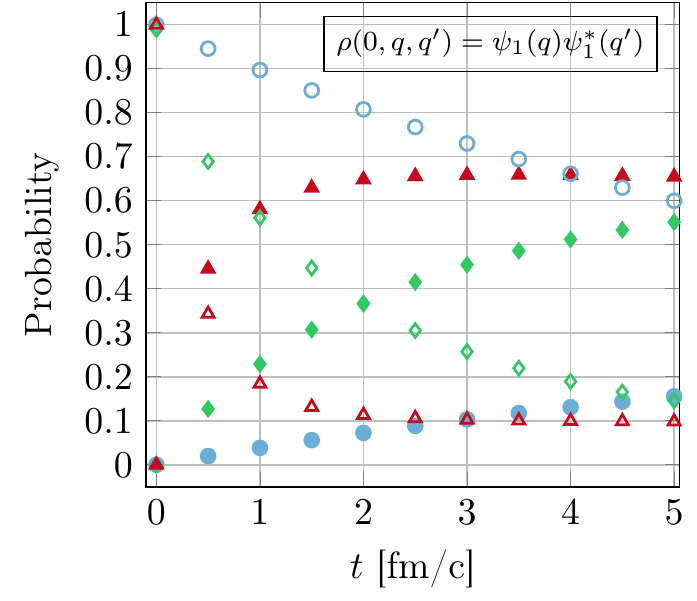}
\caption{Probabilities $P_0(t)\,,\,P_1(t)$ of having respectively the ground state $\psi_0$ and the excited state $\psi_1$ at time $t\,$. Here $l_{\psi_0}=0.08$ fm and $l_{\psi_1}=0.194$ fm, where $l_{\psi}=\sqrt{\avg{x^2}}_{\psi}$.}
\label{upsilon1}
\end{center}
\end{figure}

In this paragraph the dissociation and recombination mechanisms for two different values of heavy-quark mass are compared. This analysis is inspired by the experimental evidence that the two phenomenons are different for charmonium and bottomonium. In particular recombination effects are very small for bottomonium. Here the values $m_c= 1.2$ GeV and $m_b= 4.7$ GeV have been used for the mass of the charm and bottom quark respectively.\\
Contrasting Fig.\ref{upsilon1} with Fig.\ref{fig:probab-entropy} of section \ref{sec:numerical}, it is clear that the ground state of the heavier quarkonium is less affected by the medium than the lighter quarkonium (see bound-state probabilities on the left panel of Fig.\ref{upsilon1}, where the initial density matrix corresponds to the ground state).
\begin{figure}[t!]
\begin{center}
\includegraphics[width=7cm]{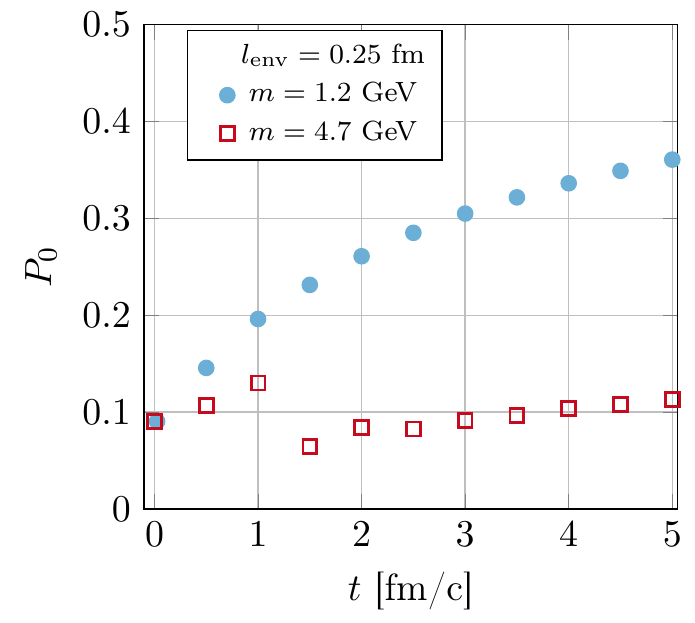}
\includegraphics[width=7cm]{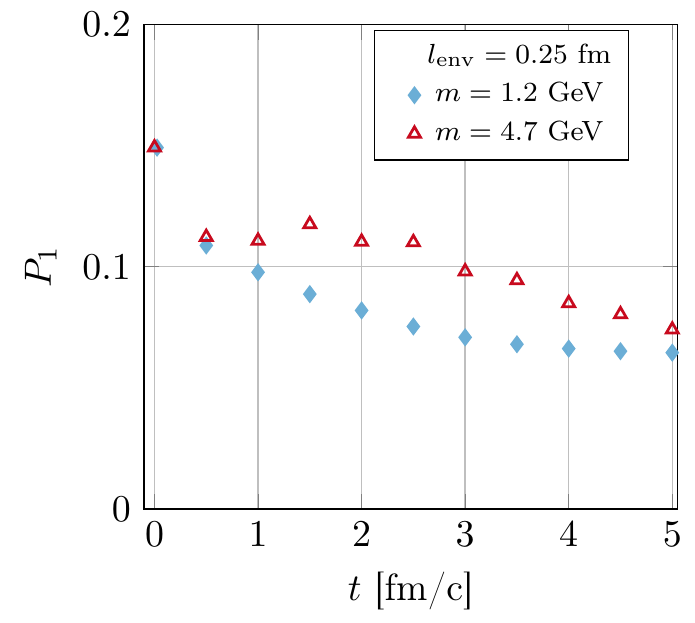}
\caption{On the left (right): Probability of having the ground (excited) state at time $t$ for different masses of the heavy quarks, starting off with an initial density matrix corresponding to a thermal scattering state. For $m=1.2$ GeV one has $\delta=2.5$ fm and $p=-0.69$ GeV, whereas $\delta=1.26$ fm and $p=-1.37$ GeV for $m=4.7$ GeV.}
\label{upsilon2}
\end{center}
\end{figure}
This can be understood from the fact that for the heavier quarkonium all the correlation lengths of the environment ($l_{\rm env}$) are greater than or equal to the size of the ground state ($l_{\psi_0}$), hence the system is mildly perturbed by the medium. The same behaviour is found when one starts off with the excited state and $l_{\rm env}>l_{\psi_1}$ (see circles and diamonds on the right panel of Fig.\ref{upsilon1}). However, when $l_{\rm env}<l_{\psi_1}$ (see triangles on the right panel of Fig.\ref{upsilon1}), the bound-state probabilities evolve in time very similarly to the ones in Fig.\ref{fig:probab-entropy} for the lighter quarkonium. In fact, in this case the dissociation process seems slightly more significant for the heavier quarkonium than for the lighter one.\\
Fig.\ref{upsilon2} shows the time evolution of the bound-state probabilities when the initial state corresponds to a thermal scattering state ($T=0.8$ GeV), as defined in eq.(\ref{eq:scatteringstate}), for a fixed $l_{\rm env}$. If on one hand the probability of getting the excited state at time $t$ (right panel) does not depend much on the mass of the heavy quark, on the other hand the probability of the scattering state of getting trapped into the ground state is largely dependent on the mass of quarkonium. In fact the left panel of Fig.\ref{upsilon2} shows that the scattering state with the heavier mass does not tend to form the ground state, whereas the one with the lighter mass clearly does.


\bibliography{refs}
\bibliographystyle{JHEP}

\end{document}